\documentclass[aps,pre,reprint,groupedaddress,floatfix]{revtex4-1}
\usepackage{amsmath}
\usepackage{amsfonts}
\usepackage{amssymb}         %additional math symbols (see amsguide.tex)
\usepackage{amsopn}
\usepackage{amsthm}
\usepackage{mathtools}
\usepackage{qcircuit}
\usepackage{hyperref}
\usepackage{multirow}
\usepackage[caption=false]{subfig}
\newcommand{\R}{\ensuremath{\mathbb{R}}}

\DeclarePairedDelimiter\bra{\langle}{\rvert}
\DeclarePairedDelimiter\ket{\lvert}{\rangle}
\DeclarePairedDelimiterX\braket[2]{\langle}{\rangle}{#1 \delimsize\vert #2}
\DeclarePairedDelimiterX\dotp[2]{\langle}{\rangle}{#1, #2}

\begin{document}

%\preprint{}

\title{Performance of hybrid quantum/classical variational
  heuristics for combinatorial optimization}


\author{Giacomo Nannicini}
\email{nannicini@us.ibm.com}
%\thanks{}
\affiliation{IBM T.J.  Watson  Research  Center,  Yorktown  Heights,  NY 10598,  USA}


\date{\today}

\begin{abstract}
  The recent literature on near-term applications for quantum
  computers contains several examples of the applications of hybrid
  quantum/classical variational approaches. This methodology can be
  applied to a variety of optimization problems, but its practical
  performance is not well studied yet. This paper moves some steps in
  the direction of characterizing the practical performance of the
  methodology, in the context of finding solutions to classical
  combinatorial optimization problems.  Our study is based on
  numerical results obtained applying several classical nonlinear
  optimization algorithms to Hamiltonians for six combinatorial
  optimization problems; the experiments are conducted via noise-free
  classical simulation of the quantum circuits implemented in
  Qiskit. We empirically verify that: (1) finding the ground state is
  harder for Hamiltonians with many Pauli terms; (2) classical global
  optimization methods are more successful than local methods due to
  their ability of avoiding the numerous local optima; (3) there does
  not seem to be a clear advantage in introducing entanglement in the
  variational form.
\end{abstract}


\maketitle

\section{Introduction}
The hybrid quantum/classical variational approach is an optimization
algorithm devised for the early generation of universal quantum
computers. Variational approaches have been applied to a variety of
fields, e.g., chemistry
\cite{peruzzo2014variational,kandala2017hardware}, machine learning
\cite{romero2017quantum,farhi2018classification}, optimization
\cite{moll2017quantum,fried2017qtorch,farhi2017quantum}.  In broad
terms, a variational approach works by choosing a parametrization of
the quantum state that depends on a relatively small set of
parameters, then using classical optimization routines to try to
determine values of the parameters corresponding to a quantum state
that maximizes or minimizes a given utility function. Typically, the
utility function is a Hamiltonian encoding the total energy of the
system, to be minimized. This directly relates to an optimization
context: the same idea can be applied to classical combinatorial
optimization problems, provided that we can construct a Hamiltonian
encoding the objective function of the optimization problem.

This report summarizes our experience in using classical
derivative-free optimization methods to try to find good solutions for
these problems, and sheds some lights on limitations that should be
overcome to increase the effectiveness of the variational
approach. All our computational experiments are based on noise-free
simulations of quantum hardware, and we therefore have access to the
full quantum state with which we can exactly evaluate the
Hamiltonians. The conclusions of the study might be considerably
different if real hardware had been used, considering the inherent
amount of noise that affects the computations. To the best of our
knowledge, this is the most comprehensive numerical study of a hybrid
quantum/classical variational method for combinatorial
optimization. The most notable limitation of our study is the fact
that we use only one type of variational form, and we do not test any
problem-dependent variational form as advocated by some other works
\cite{farhi2014quantum,farhi2016quantum,farhi2017quantum}.

Our study leads to the following observations. Even if the variational
form used in our experiments is guaranteed to span the ground state,
the resulting optimization problems are difficult for classical local
optimization methods, e.g., gradient descent: these methods often get
stuck in local minima that may be very far from the optimal
solution. Global optimization seems to be more reliable, but the
performance of all algorithms is very problem-dependent. In
particular, the variational approach has difficulties on the problem
classes that also appear to be the hardest for a classical
Branch-and-Bound solver. This difficulty is likely related to the
concept of ``density'' of the problem representation. Indeed, the
performance of the classical Branch-and-Bound solver can be explained
by the fact that the classical representation of these difficult
problems as binary quadratic optimization problems leads to dense
matrices, which are known to be harder to deal with than sparse
matrices. Similarly, for the variational approach our experiments show
that the number of distinct eigenvalues of the Hamiltonian is a good
indicator of the difficulty of a problem instance; this can be
explained in light of the fact that eigenpairs represent stationary
points of the optimization problem. The number of distinct eigenvalues
is often related to the number of terms in the representation of the
Hamiltonian as a weighted sum of Paulis, depending on the
weights. This observation may provide an easy way to quickly estimate
the difficulty of finding the ground state with the variational
approach on a given Hamiltonian. Furthermore, for these classes of
problems, which yield diagonal Hamiltonians, it is unclear if
two-qubit entangling gates help accelerate convergence to the ground
state. Finally, regardless of the method used and the problem class,
attaining a good approximation ratio or attaining a good probability
of sampling the optimal solution seems to require a large number of
iterations of the classical optimization routine; the scaling of the
performance with respect to problem size (limited to what can be
ascertained in a study that considers at most 18 qubits) indicates
that as problem size increases, the number of necessary iterations
grows more than linearly, which is expected when dealing with
nonconvex problems. Considering the crucial role played by the
variational form in this type of method, it is important that future
research efforts carefully consider the choice of variational form and
its effect on the performance of the optimization algorithm.

\section{The variational approach}
The hybrid quantum/classical variational approach aims to find the
quantum state attaining minimum energy for a given Hamiltonian by
varying a set of parameters that control the quantum state. The
algorithm that varies the parameters is a classical optimization
algorithm. Formally, let $H$ be the Hamiltonian encoding the total
energy of a system, let $\theta$ be a vector of parameters, and let
$\ket{\psi(\theta)} = U(\theta) \ket{0}$ be the quantum state obtained
by applying a given parametrized quantum circuit $U(\theta)$ to the
initial state $\ket{0}$; for example, the quantum circuit $U(\theta)$
could include some rotations, and the vector $\theta$ encodes the
rotation angles. The variational approach aims to determine:
\begin{equation}
  \label{eq:vqe}
  \min_{\theta} \bra{\psi(\theta)} H \ket{\psi(\theta)}.
\end{equation}
It is well known that since $\ket{\psi(\theta)}$ is normalized, the
minimum value of \eqref{eq:vqe} is bounded below by the minimum
eigenvalue $\lambda_{\min}$ of $H$, and in fact $\lambda_{\min} =
\min_{\ket{\psi}} \bra{\psi} H \ket{\psi}$. Determining such minimum
value is in general NP-hard, as will be shown in the next section by
encoding several NP-hard problems into this framework. The
optimization of \eqref{eq:vqe} can be performed in a hybrid setting
that uses a classical computer running an iterative algorithm to
select $\theta$, and a quantum computer to compute information about
$\bra{\psi(\theta)} H \ket{\psi(\theta)}$ for given $\theta$, for
example the value of $\bra{\psi(\theta)} H \ket{\psi(\theta)}$ itself
or its derivatives with respect to $\theta$. Since finding the minimum
of \eqref{eq:vqe} is an approximation of the problem of finding the
minimum eigenvalue of $H$, this approach is typically called the
variational quantum eigensolver (VQE) in the literature
\cite{mcclean2016theory}.

The unitary matrix $U(\theta)$ is typically called the variational
form or ansatz. Clearly the choice of the variational form has a
fundamental role. In certain settings, it is possible to show that a
specific variational form spans the optimal solution to a class of
problems, or that there exist efficient algorithms to optimize
$\theta$ under some conditions \cite{farhi2014quantum}. However,
appropriately choosing $U(\theta)$ is in general a difficult task.
Determining an appropriate classical algorithm to optimize over
$\theta$ is also difficult in general. Existing works in the
literature typically employ iterative continuous optimization
algorithm, e.g., various forms of gradient descent or direct search
methods \cite{romero2018strategies,kandala2017hardware}.

\section{Hamiltonians for binary optimization problems}
\label{s:problems}
The most natural formulation of combinatorial optimization problems on
a quantum computer is via an Ising spin glass model, which directly
translates into a Hamiltonian. Indeed, we have:
\begin{equation*}
  Z = \begin{pmatrix} 1 & 0 \\ 0 &-1 \end{pmatrix} \qquad Z\ket{0} =
  \ket{0} \qquad Z\ket{1} = -\ket{1},
\end{equation*}
and the two eigenvalues $\pm 1$ of $Z$ correspond to the positive and
negative spin. The Ising spin glass model can be seen as a quadratic
unconstrained binary optimization problem, and it inherits its
hardness \cite{barahona1982computational}. In general, computing the
partition function of an Ising model is NP-complete
\cite{istrail2000statistical}. Because of this, any problem in NP can
be reduced to an Ising model; we are particularly interested in
combinatorial problems that have a natural mapping to Ising spin glass
models, i.e., problems with the property that if the original instance
has size $n$, we need only $n$ qubits for an Ising spin glass
representation.

As we will see in the following, some problems are naturally
formulated in terms of spin variables $\pm 1$, whereas others have a
more natural formulation in terms of 0-1 variables. The transformation
between the two types of variables is straightforward, see e.g.,
\cite{lucas2014ising}~. The main idea is as follows. Consider a binary
quadratic unconstrained optimization problem:
\begin{equation}
  \label{eq:qp}
  \min \{ c^{\top} x + x^{\top} Q x : x \in \{0,1\}^n \},
\end{equation}
then transform \eqref{eq:qp} into an Ising model using the
substitution $x_j = \frac{y^Z_j + 1}{2}$, where $x_j \in \{0,1\}$ and
$y^Z_j \in \{-1,1\}$ for $j=1,\dots,n$; we use the superscript $Z$ to
distinguish $\pm 1$ spins from 0-1 variables. Since \eqref{eq:qp} is a
quadratic model, the substitution yields a summation of terms, each of
which contains either one or two $y^Z_j$ variables. The Hamiltonian is
then a summation of weighted tensor products of $Z$ Pauli operators,
where each term of the summation contains at most two $Z$s.
%% It should be kept in mind that a measurement of $0$ on the $j$-th
%% qubit corresponds to $x_j = 1$, and a measurement of $1$ corresponds
%% to $x_j = 0$.
Furthermore, since $Z$ is diagonal, the Hamiltonian resulting
from this transformation is diagonal.

If the original binary quadratic optimization problem is constrained,
the approach mentioned above can still be applied by adding
appropriate (quadratic) penalties for constraint violations in the
objective function. In the cases of relevance for this paper, the
additional constraints for \eqref{eq:qp} can be expressed as the
requirement $Ax = b$ for some choice of $A, b$; in this case, it is
sufficient to add the term $\alpha \|Ax - b\|^2$ to the objective
function \eqref{eq:qp} with a sufficiently large $\alpha$, to ensure
that the unconstrained formulation has the same optimum as the
original constrained formulation.

We now describe the six classes of combinatorial optimization problem
employed in our numerical study.

\subsection{Maximum stable set}
Given an undirected graph $G = (V,E)$, a stable set (also called
independent set) is a set of mutually nonadjacent vertices. We are
interested in determining a stable set of maximum cardinality: we
label this problem {\sc StableSet}. Assuming that $V = \{1,\dots,n\}$,
the problem can be formulated as:
\begin{equation*}
  \max \left\{ \sum_{j \in V} x_j - \sum_{(i,j) \in E} x_i x_j : x \in
  \{0,1\}^n\right\}.
\end{equation*}
Indeed, the first summation in the objective function represents the
cardinality of the stable set, while the second part penalizes
including two adjacent vertices. It is straightforward to check that
the penalty always offsets the objective function increase derived
from selecting a vertex that is adjacent to an already selected
vertex. This problem is a specific case of the set packing problem
\cite{lucas2014ising}. It is one of six basic NP-complete problems
discussed in the seminal work of
\cite{garey1972computers}~. Transforming the 0-1 binary variables into
$\pm 1$ spins gives the Hamiltonian.

\subsection{Maximum 3-satisfiability}
The maximum 3-satisfability problem, {\sc Max3SAT} in the following,
tries to determine an assignment of Boolean variables that satisfies
the largest number of clauses of a Boolean formula in conjunctive
normal form, where each clause has exactly three literals. This is one
of the six basic NP-complete problems in \cite{garey1972computers}~.
There are several approaches to construct a Hamiltonian for {\sc
  Max3SAT}. The approach followed in this paper is to transform an
instance of {\sc Max3SAT} with $m$ clauses into an instance of {\sc
  StableSet} on a suitably constructed graph with $3m$ vertices. This
transformation is well-known, and we refer the reader to
\cite{garey1972computers,lucas2014ising} for details. We remark that
in principle we can model {\sc Max3SAT} with $n$ Boolean variables
using $n$ Ising spins and a $3$-local Hamiltonian, i.e., a tensor
product of three Pauli terms for each clause. However, we employ the
transformation to {\sc StableSet} for two reasons: first, it allows to
study the behavior of VQE on random instances of {\sc StableSet}
versus structured instances of the same problem; second, the
formulation with products of three Pauli terms cannot be directly
translated into a quadratic unconstrained model with $n$ 0-1
variables, and we use this direct transformation in Section
\ref{s:cplex} when assessing the difficulty of these problem instances
with a classical Branch-and-Bound solver.

\subsection{Number partitioning}
Given a set of numbers $S := \{a_1,\dots,a_n\}$, the problem of number
partitioning ({\sc Partition}) asks to determine $P_1,P_2 \subset
\{1,\dots,n\}, P_1 \cup P_2 = \{1,\dots,n\}, P_1 \cap P_2 = \emptyset$
such that $|\sum_{j \in P_1} a_j - \sum_{j \in P_2} a_j|$ is
minimum. To construct a Hamiltonian for this problem, notice that if
we associate a Ising spin variable $y^Z_j \in \{-1,1\}$ to each
number $a_1,\dots,a_n$, we have $\sum_{j=1,\dots,n} y^Z_j =
\sum_{j : y^Z_j = 1} a_j - \sum_{j : y^Z_j = -1}
a_j$. Furthermore, minimizing $|\sum_{j \in P_1} a_j - \sum_{j \in
  P_2} a_j|$ is equivalent to minimizing $|\sum_{j \in P_1} a_j -
\sum_{j \in P_2} a_j|^2$, and we can thus write:
\begin{equation*}
  \min \left\{ \left(\sum_{j=1,\dots,n} y^Z_j\right)^2 :
  y^Z_j \in \{-1,1\} \right\}.
\end{equation*}
Expanding the square gives the Hamiltonian in the desired form. {\sc
  Partition} is one of the six basic NP-complete problems in
\cite{garey1972computers}~.

\subsection{Maximum cut}
Given an undirected graph $G = (V,E)$ with weights $w_{ij}$ on the
edges, the maximum cut problem ({\sc MaxCut}) calls for determining a
partition of $V$ into disjoint sets $V_1, V_2$ such that 
\begin{equation*}
  \sum_{\substack{(i,j) \in E \\ i \in V_1, j \in V_2}} w_{ij}
\end{equation*}
is maximum, i.e., the sum of the weights of edges with endpoints on
opposite sides of the partition. An Ising spin glass model without
field is essentially a weighted {\sc MaxCut} problem
\cite{barahona1982computational}. The problem can be formulated as:
\begin{equation*}
  \max \left\{ \sum_{(i,j) \in E} w_{ij} y^Z_i y^Z_j -
  \sum_{(i,j) \in E} \frac{w_{ij}}{2} \right\}.
\end{equation*}

\subsection{Market split}
The market split problem \cite{marketsplitjournal} can be described as
the problem of assigning the $n$ customers of a firm that sells $m$
products to two subdivisions of the same firm, in such a way that the
two subdivisions retain roughly an equal share of the
market. Formally, we are given a matrix $A$ with nonnegative entries
$a_{ij}$ that represent the amount of product $i$ bought by customer
$j$. We want to determine a 0-1 assignment $x_j$ for each customer $j$
so that for each product $i$, $\sum_{j=1}^n a_{ij} x_j \approx
\sum_{j=1}^n a_{ij}$. If we let $b$ be the vector with entries $b_i =
\left\lfloor \sum_{j=1}^n a_{ij} \right\rfloor$, then this is simply
the problem:
\begin{equation*}
  \min \left\{ \|Ax - b\|^2 : x \in \{0,1\}^n \right\}.
\end{equation*}
Epanding the square and transforming the 0-1 binary variables into
$\pm 1$ spins gives the Hamiltonian. This problem is known to be very
difficult for classical algorithms based on Branch-and-Bound
\cite{aardalmarkshare}.

\subsection{Traveling salesman problem}
Given an undirected complete graph $G = (V,E)$ with weights $w_{ij}$
on the edges, the traveling salesman problem ({\sc TSP}) aims to find
a Hamiltonian cycle of minimum weight, i.e., a cycle that visits all
nodes of the graph and such that the sum of the edge weights is
minimum. To formulate this problem we use the formulation given in
\cite{qiskit}~. Let $n$ be the number of nodes. For $i,p=1,\dots,n$,
let $x_{i,p}$ be $1$ if node $i$ appears in position $p$ in the cycle,
0 otherwise. Fixing the first node of the cycle to be the node with
index label $1$, i.e., $x_{1,1} = 1$, {\sc TSP} can be formulated as:
\begin{equation*}
  \left.
  \begin{array}{rrcl}
    \min & \sum_{i,j=1}^{n-1} w_{ij} \sum_{p=1}^{n-1} x_{i,p} x_{j,p+1} + \\
    & \sum_{j=1}^{n-1} w_{j1} x_{j,n} & & \\
    \forall i=1,\dots,n & \sum_{p=1}^n x_{i,p} &=& 1 \\
    \forall p=1,\dots,n & \sum_{i=1}^n x_{i,p} &=& 1 \\
    & x_{1,1} &=& 1 \\
    \forall i,p=1,\dots,n & x &\in& \{0,1\}.
  \end{array}
  \right\}
\end{equation*}
To derive a Hamiltonian for this problem, we penalize the violation of
the constraints in the objective function inserting terms of the form
$\alpha(\sum_{p=1}^n x_{i,p} - 1)^2$, where $\alpha$ is sufficiently
large, e.g., $\alpha = n\max_{(i,j) \in E} w_{ij}$. {\sc TSP} (rather,
Hamiltonian cycle) is one of six basic NP-complete problems in
\cite{garey1972computers}~. Note that the formulation requires $n^2$
binary variables, so the number of qubits does not scale linearly in
the problem size; this is the only problem in our test set with this
property.

\section{Data and experimental setup}
\label{s:exp}
This section describes the procedure used to generate random instances
of each class, as well as the overall setup used for our experiments.

\subsection{Generation of random instances}
Given the desired number of qubits $q$, we generate instances as
follows:
\begin{itemize}
\item {\sc StableSet}: we generate a random Erd\H{o}s-R\'enyi graph
  with $q$ nodes and edge probability 0.3.
\item {\sc Max3SAT}: we generate a random formula in conjunctive
  normal form with $\left\lfloor q/3 \right\rfloor$ Boolean variables
  and $\left\lfloor q/2 \right\rfloor$ clauses. Clauses are generated
  sequentially, adding one literal (positive or negative) chosen
  uniformly at random among literals that do not appear in the same
  clause.
\item {\sc Partition}: we generate $q$ integers in the interval
  $[1,q^2+1]$, chosen uniformly at random.
\item {\sc MaxCut}: we generate a complete graph with $q$ nodes and
  integer weights selected uniformly at random in the interval
  $[-10,10]$.
\item {\sc Marketsplit}: we follow the procedure described in
  \cite{marketsplitjournal}~, for $q$ binary variables.
\item {\sc TSP}: we generate a complete graph with $\sqrt{q}+1$ nodes
  and integer weights selected uniformly at random in the interval
  $[0,9]$. Note that the resulting problem instance is in general not
  symmetric.
\end{itemize}

\subsection{Experimental setup}
\begin{figure}[tbp]
  \subfloat[Instances between 6 and 12 qubits.]{
    \includegraphics[width=0.5\textwidth]{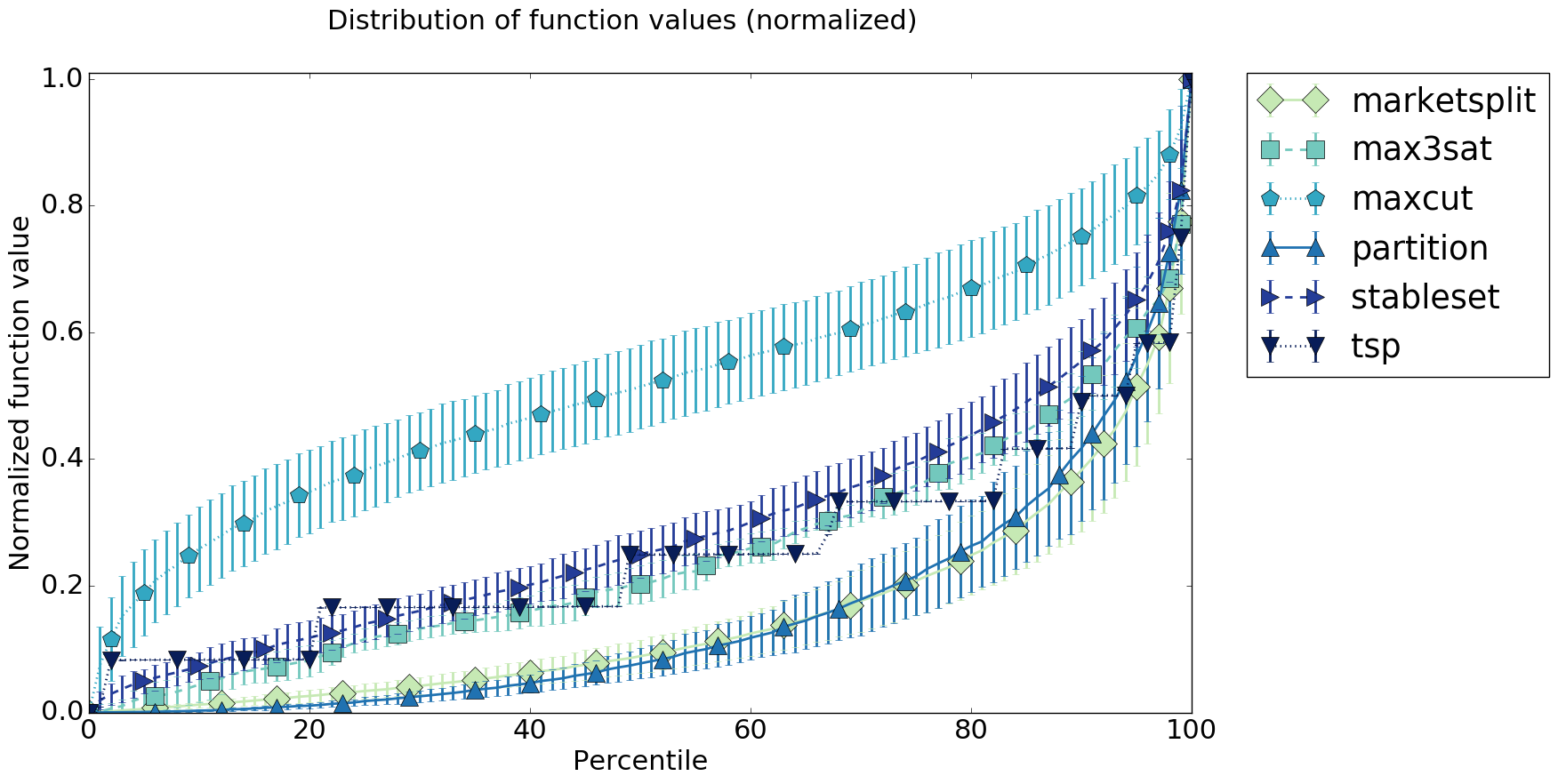}
  }
  
  \subfloat[Instances between 13 and 18 qubits.]{
    \includegraphics[width=0.5\textwidth]{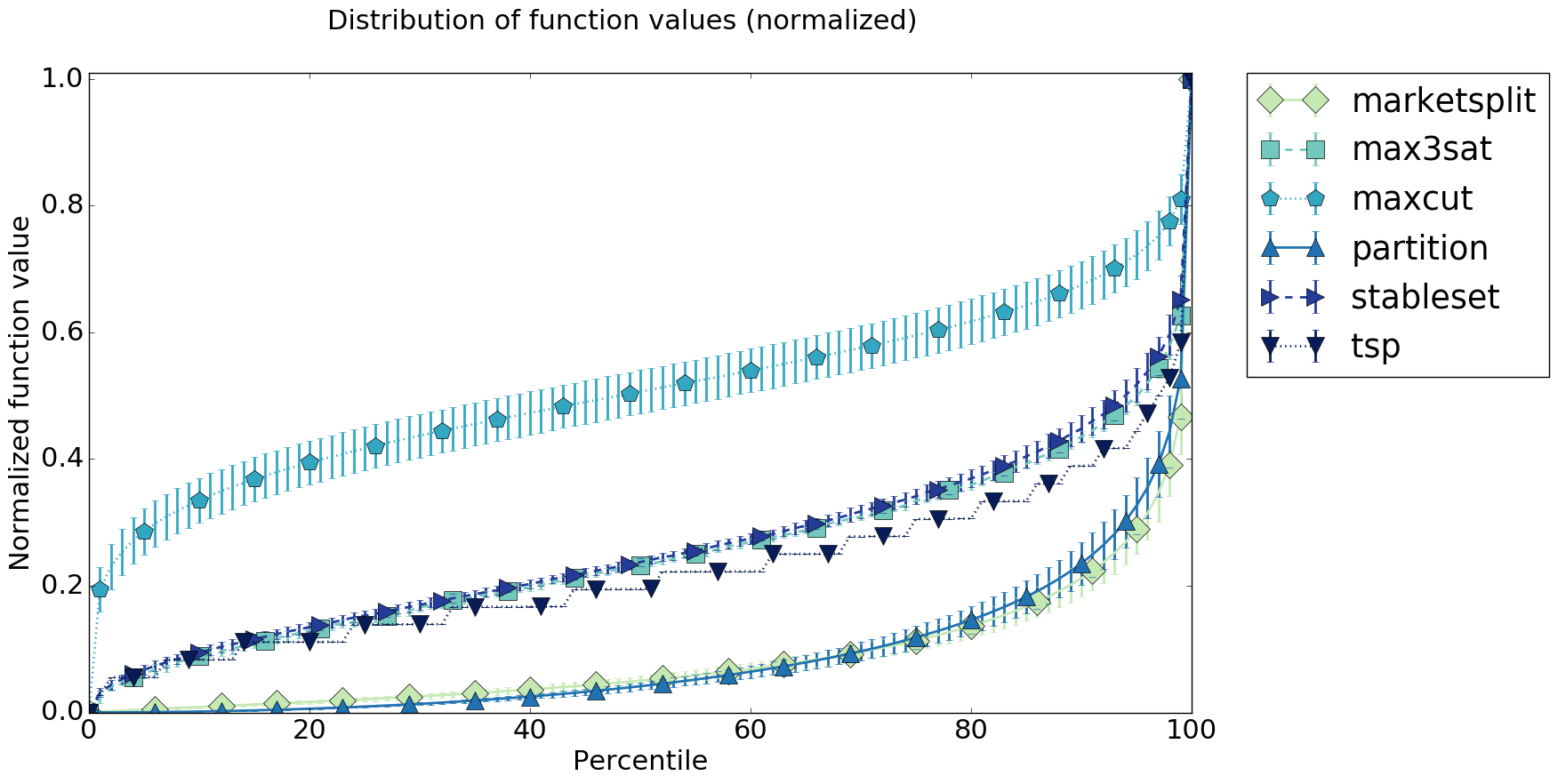}
  }
  \caption{Distribution of the objective function value of all
    feasible solutions for the random instances used in the numerical
    experiments, averaged across all instances of a certain size. The
    $y$ axis value is the approximation ratio. Error bars extend
    between $\pm$ the standard deviation.}
  \label{fig:value_dist}
\end{figure}

We applied the VQE to all problems described in the previous
subsection, testing all problem sizes from 6 to 18 qubits; because of
the Hamiltonian formulation used, Max3SAT requires the number of
qubits to be a multiple of 3, and TSP requires the number of qubits to
be a perfect square, hence these two problems were only tested for
sizes in the range $[6, 18]$ that satisfy the stated restrictions. For
each problem type and size, we repeat the experiment 20 times with a
different random seed; the random seed affects the instance itself (as
it is randomly generated) and the starting point of the optimization
algorithm. Note that we provide the same sequence of random seeds to
all optimization algorithms, so that all optimization algorithms solve
the same sequence of problems with the same sequence of starting
points. The average number of terms in the Hamiltonian for some
problem sizes, as well as the number of distinct eigenvalues, are
reported in Table \ref{tab:terms}, where we can see that a large
number of terms does not necessarily correspond to more distinct
eigenvalues, because of integer weights. The distribution of the
objective function value of the feasible solutions for the randomly
generated instances is plotted in Fig.~\ref{fig:value_dist}. It shows
that {\sc Partition} and {\sc Marketsplit} instances have many
solutions that are close to the optimum in relative terms, {\sc
  MaxCut} has very few, and the other problem classes are somewhere in
between.

\begin{table}[tb]
  \centering
  {\scriptsize
  \begin{tabular}{|l|c|c|c|c|c|c|c|c|}
    \hline
    & \multicolumn{4}{c|}{\# terms} & \multicolumn{4}{c|}{\# distinct eig.} \\
    \hline
    $q$ & 15 & 16 & 17 & 18 & 15 & 16 & 17 & 18\\
    \hline
        {\sc MaxCut} & 99 & 113 & 129 & 145 & 201 & 225 & 251 & 228 \\
        {\sc TSP} & -- & 99 & -- & --  & -- & 5291 & -- & -- \\
        {\sc Max3SAT} & 37 & -- & -- & 43 & 65 & -- & -- & 78 \\
        {\sc Partition} & 105 & 120 & 136 & 153 & 710 & 883 & 1081 & 1304 \\
        {\sc Marketsplit} & 120 & 136 & 153 & 171 & 22204 & 37564 & 57532 & 81151 \\
        {\sc Stableset} & 48 & 53 & 60 & 65 & 107 & 120 & 138 & 150 \\
        \hline
  \end{tabular}
  }
  \caption{Average number of terms in the Hamiltonian (each term is a
    tensor product of Pauli $Z$), and average number of distinct
    eigenvalues. All numbers are rounded to the nearest integer.}
  \label{tab:terms}
\end{table}

%% All results reported in the following are based on median values over
%% these 20 random seeds: each problem type and size contributes only one
%% datapoint (as opposed to 20), but taking the median over 20 instances
%% increases the robustness of our analysis to randomness.
To evaluate the progress of VQE toward reaching an optimal solution
for the problem at hand, we employ a methodology based on the data
profiles described in \cite{morewild09}~. More details regarding the
meaning of each graph are given in the corresponding subsections. The
optimum for each problem is computed using the classical solver IBM
ILOG Cplex 12.7.1, which can certify optimality; we use an optimal 0-1
solution obtained by Cplex to compute the value of the ground state of
the corresponding Hamiltonian.

We tested five optimization algorithms:
\begin{enumerate}
\item Limited-memory BFGS (LBFGS) \cite{byrd1995limited}: a
  quasi-Newton local optimization method. We use the SciPy
  implementation of this algorithm, in which the gradient is estimated
  numerically by finite differences.
\item Constrained Optimization By Linear Approximation (COBYLA)
  \cite{powell94direct}: a model-based local optimization method that
  builds a linear approximation of the objective function over a
  simplex. We use the original FORTRAN implementation through its
  SciPy interface. 
\item RBFOpt \cite{nannirbfopt}: a model-based global search method
  that builds an adaptive radial basis function interpolant of the
  objective function. The algorithm alternates between a global search
  and a local search that follows a trust region framework
  \cite{connbook}.
\item Modified Powell's conjugate direction method (PCD)
  \cite{powell64efficient}: a pattern search local optimization method
  that searches along a given set of directions, which is updated at
  every iteration. We use the implementation in SciPy.
\item Simultaneous Perturbation Stochastic Approximation (SPSA)
  \cite{spall92multivariate}: a model-based local optimization method
  that builds a gradient approximation using two function evaluations
  per iteration. We use the implementation found in QISKit
  \cite{qiskit}.
\end{enumerate}
In the brief description above, we classify as ``local'' algorithms
those which converge to a (at least) first-order stationary point, and
``global'' algorithms those that do not employ convergence criteria
based on first-order stationarity, but rather have a mechanism to
escape any local optimum. In our tests, the performance of LBFGS and
COBYLA is very similar across the board.  PCD and SPSA performed
considerably worse than the remaining algorithms in our
experiments. This is not surprising: PCD was not designed to be
parsimonious in the number of function evaluations, and therefore
exhibits slower convergence; whereas SPSA was designed to be robust to
noise, but this robustness comes at a price and is not exploited at
all in our noise-free setting. Overall, numerical experiments using
PCD and SPSA are simply slower and do not yield further insight with
respect to looking at the first three optimization algorithms
alone. Other popular algorithms such as Nelder-Mead and genetic
algorithms were not considered as they usually yield inferior results
in mathematical benchmarks \cite{connbook}.  For these reasons, in the
following we only report results for COBYLA, RBFOpt and occasionally
LBFGS.
%; additional graphs for LBFGS are given in the Appendix.

\subsection{Analysis of difficulty with classical Branch-and-Bound}
\label{s:cplex}
\begin{figure}[tbp]
  \centering
  \includegraphics[width=0.5\textwidth]{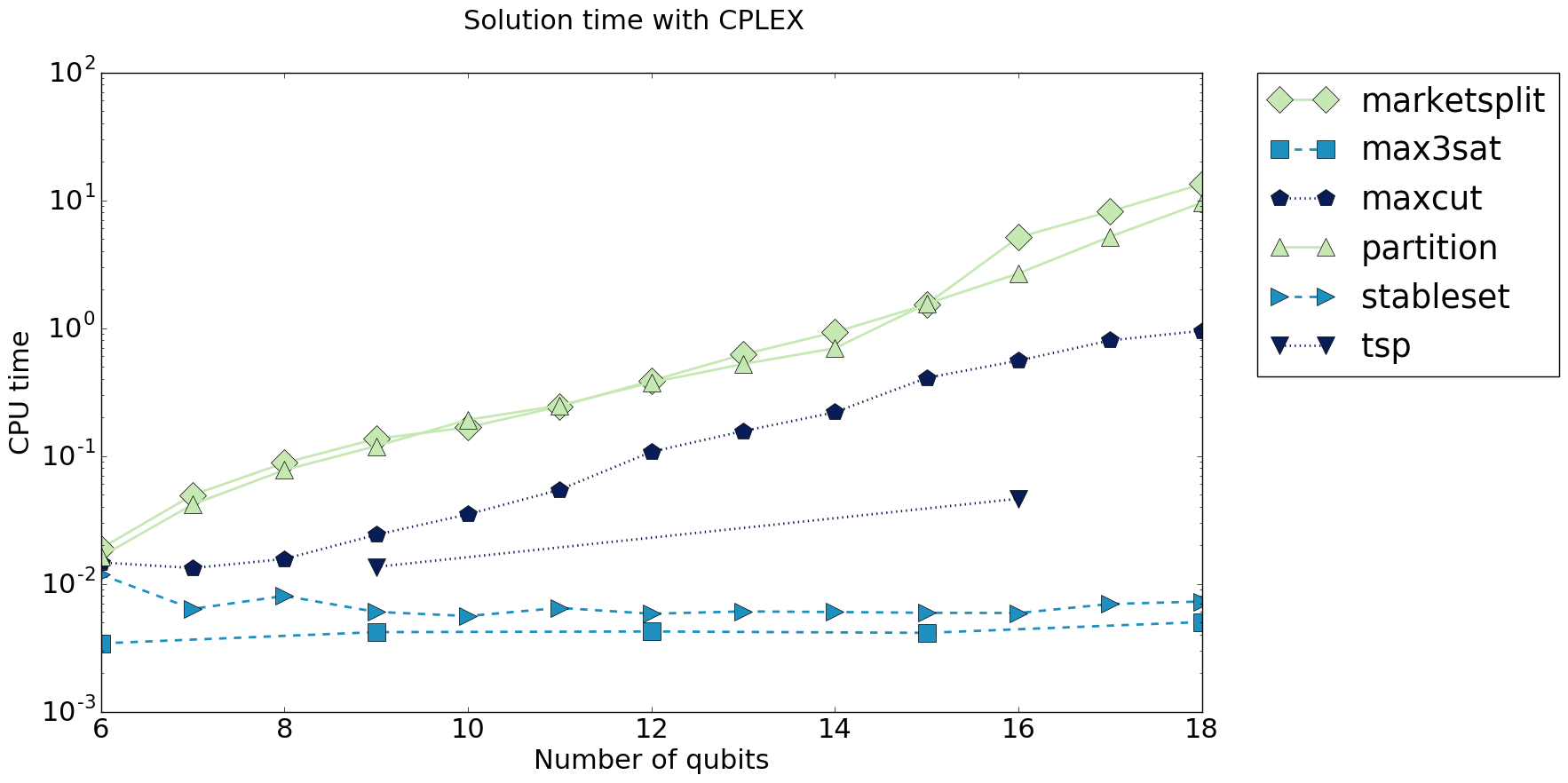}
  \caption{Average solution times using IBM ILOG Cplex,
    single-threaded. The $y$ axis is on a log scale.}
  \label{fig:cplex_times}
\end{figure}

To quickly assess the difficulty of instances in our test set on
classical computers, we transformed each Hamiltonian into a quadratic
unconstrained binary optimization problem, and we solved it to global
optimality using using the Branch-and-Bound algorithm in the
commercial integer programming solver IBM ILOG Cplex 12.7.1. The
average solution times are given in Fig.~\ref{fig:cplex_times}. The
graph shows that running times for {\sc Partition}, {\sc Marketsplit}
and {\sc Maxcut} scale exponentially with problem size, while the
other problem classes appear considerabily easier. 

We remark that we did not make any attempt at fine-tuning the
algorithm or improving the problem formulation: we directly translated
the Hamiltonian into a quadratic form with binary variables, and let
Cplex solve the instance with its default parameters \footnote{Because
  the quadratic form is in general nonconvex, to solve the instance to
  global optimality we set the parameter {\tt optimalitytarget} to
  {\tt optimalglobal}, we set {\tt absmipgap} to 0.1 and {\tt mipgap}
  to 0.}. It is well known that the performance of integer programming
models depends heavily on the particular formulation of the
optimization problem; our automatic translation of the Hamiltonian
yields very weak formulations, therefore in practice one can expect
significant improvements (potentially orders of magnitude) using
better classical methodologies. In other words,
Fig.~\ref{fig:cplex_times} is far from representing the state of the
art of classical optimization for the problem classes under
consideration.

To understand what causes the difficulty of certain problem classes
for IBM ILOG Cplex, we looked at two properties that are known to
affect the performance of solution methods for quadratic optimization
problems, namely: the density of $Q$, and the distribution of its
eigenvalues. The linear term $c$ of the cost function was added to the
diagonal of $Q$, since for binary variables $x^2 = x$. Statistics are
reported in Table \ref{tab:density}. These data suggest that density
strongly correlates with problem difficulty: most instances have many
negative eigenvalues (all problems are expressed as minimization
problems), but only some instances are dense ({\sc Marketsplit} and
{\sc Partition} are fully dense, followed by {\sc MaxCut}), and these
appear to be the hardest to solve for Cplex.

\begin{table}[tb]
  \centering
  \begin{tabular}{|l|c|c|c|c|c|c|c|c|}
    \hline
    & \multicolumn{4}{c|}{Density} & \multicolumn{4}{c|}{Negative eigenvalues} \\
    \hline
    $q$ & 15 & 16 & 17 & 18 & 15 & 16 & 17 & 18\\
    \hline
        {\sc MaxCut} & 0.95 & 0.95 & 0.95 & 0.95 & 0.62 & 0.61 & 0.59 & 0.60 \\
        {\sc TSP} & -- & 0.72 & -- & -- & -- & 1.00 & -- & -- \\
        {\sc Max3SAT} & 0.31 & -- & -- & 0.25 & 1.00 & -- & -- & 1.00 \\
        {\sc Partition} & 1.00 & 1.00 & 1.00 & 1.00 & 1.00 & 1.00 & 1.00 & 1.00 \\
        {\sc Marketsplit} & 1.00 & 1.00 & 1.00 & 1.00 & 1.00 & 1.00 & 1.00 & 1.00 \\
        {\sc Stableset} & 0.40 & 0.39 & 0.39 & 0.38 & 1.00 & 1.00 & 1.00 & 1.00 \\
        \hline
  \end{tabular}

  \caption{Average density (expressed as the number of nonzero
    elements in $Q$ over the total number of elements) and fraction of
    negative eigenvalues for the test instances.}
  \label{tab:density}
\end{table}

\subsection{Variational form}
\label{s:varform}
The choice of the variational form is crucial for the performance of
VQE; in particular, if the ground state (minimum energy state) of the
Hamiltonian cannot be attained by a given variational form, then VQE
will never reach the optimum energy. For combinatorial optimization
problems it is easy to construct variational forms that are guaranteed
to contain the ground state in their span: it is sufficient to ensure
that the variational form can generate any binary string on $q$
qubits.

\begin{figure}
 \leavevmode
 \centering
 \Qcircuit @C=1em @R=.6em {
   & \qw & \gate{Y(\theta_1)} & \ctrl{1} & \qw      & \ctrl{2} & \gate{Y(\theta_4)} & \ctrl{1} & \qw      & \ctrl{2} & \gate{Y(\theta_7)} & \qw\\
   & \qw & \gate{Y(\theta_2)} & \gate{Z} & \ctrl{1} & \qw      & \gate{Y(\theta_5)} & \gate{Z} & \ctrl{1} & \qw      & \gate{Y(\theta_8)} & \qw\\
   & \qw & \gate{Y(\theta_3)} & \qw      & \gate{Z} & \gate{Z} & \gate{Y(\theta_6)} &   \qw    & \gate{Z} & \gate{Z} & \gate{Y(\theta_9)} & \qw
   \gategroup{1}{3}{3}{3}{1.3em}{--} \gategroup{1}{4}{3}{7}{1.3em}{--}
   \gategroup{1}{8}{3}{11}{1.3em}{--}
 }
  \caption{Example of the variational form on three qubits. Each box represents a layer.}
  \label{fig:var_circuit}
\end{figure}
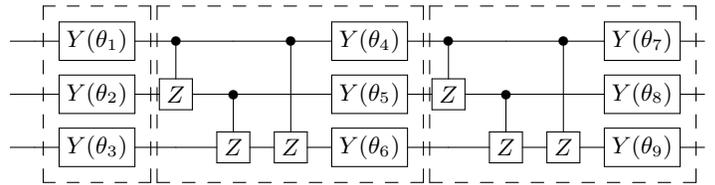

In our experiments, we use a variational form constructed in layers,
see \cite{kandala2017hardware}. The first layer always consists in
single-qubit $Y$ rotations, with one variational parameter per qubit
to determine the rotation angle. This ensures that any binary string
can be obtained with just the first layer. Each additional layer after
the first contains entangling gates, more specifically controlled-$Z$
gates applied to all qubit pairs, followed by another set of
single-qubit $Y$ rotations with one variational parameter each to
represent the angle. Thus, each layer has $q$ variational
parameters. The resulting circuit is exemplified in
Fig.~\ref{fig:var_circuit} for three qubits. We experimented with
nearest-neighbor controlled-$Z$ gates as well, i.e., between qubit $j$
and $j+1$ for all $j=1,\dots,q$, but this does not change the
conclusions of the study.

%% Notice that we use
%% all-to-all entanglement, since a circuit of this form with
%% nearest-neighbor entanglement only can be classically simulated in
%% polynomial time using known results on matchgates
%% \cite{van2010quantum}.

Our experiments use up to three layers of this variational form, using
the labels 1L, 2L, 3L to indicate how many have been used. For the
global classical optimization algorithm RBFOpt only 1L and 2L are
tested, because the optimization for 3L is too time consuming. By
construction, the variational form has low depth, as is the case for
ans\"atze that have been implemented in hardware
\cite{kandala2017hardware}.

The specific choice of variational form used in this paper is
justified by the fact that it is guaranteed to span the ground
state. The entangling layers, coupled with Y rotations, allow us to
control the level of entanglement, and experimentally verify whether
entanglement helps speed up convergence, see Section
\ref{s:entnoent}. We remark that for 0-1 optimization problems there
exists an optimal solution that is a computational basis state, hence
entanglement is not necessary in principle, unlike, e.g., certain
quantum chemistry problems in which the ground state is known to be an
entangled state.

\section{Results and analysis}
\label{s:results}
We now report a summary of our findings, based on plots provided in
this paper as well as further analysis not reported for space reasons.

\subsection{Convergence versus number of iterations}
\label{s:conviter}
In the first set of graphs we study the convergence of each
optimization algorithm as the number of iterations progresses, over
the entire set of problem instances. Here and in the rest of the
paper, convergence is defined in the following way. Let $\tilde{x}$ be
the initial point given to the optimization algorithm \footnote{This
  assume that the algorithm accepts and uses an initial point, e.g.,
  to start a gradient descent; all optimization algorithms tested in
  this paper accept an initial point.}, and let $x^*$ be an optimal
solution to the problem. Calling $f$ the objective function, we say
that an optimization algorithm converges to a precision of $\tau \in
[0,1]$ if it determines a point $x$ such that $\frac{f(x) -
  f(x^*)}{f(\tilde{x}) - f(x^*)} \ge 1 - \tau$. Notice that when $\tau
= 0$ this implies determining an optimal solution, whereas $\tau = 1$
is trivially satisfied by any point $x$ returned by the optimization
algorithm.

To account for different problem sizes and the dimension of the search
space, we normalize the iteration number by reporting the ``equivalent
gradient iterations'', where each gradient iteration performs $n+1$
function evaluations and $n$ is the total number of parameters of the
variational form that are being optimized. Using the variational form
described in Section \ref{s:varform}, a problem instance on $q$ qubits
with a variational form with $\ell$ layers has $n = q\ell$
parameters. We remark that $n+1$ corresponds exactly to the number of
function evaluations that are perfomed by a gradient-based method that
estimates the gradient by finite differences along the coordinate axes
(e.g., the LBFGS implementation used in our tests); for such methods,
the normalization gives an accurate count of the major iterations of
the optimization algorithm. Other methods, however, do not try to
estimate the gradient at every major iteration, but we apply the same
normalization in order to have a fair comparison. This normalization
is also standard in the derivative-free optimization literature to
account for varying problem sizes \cite{morewild09}. The maximum
number of function evaluations is set to $100(n+1)$ for all
optimization algorithms, in these and in all subsequent experiments.

\begin{figure}[tb]
  \subfloat[Constrained Optimization By Linear Approximation (COBYLA).]{
    \includegraphics[width=0.5\textwidth]{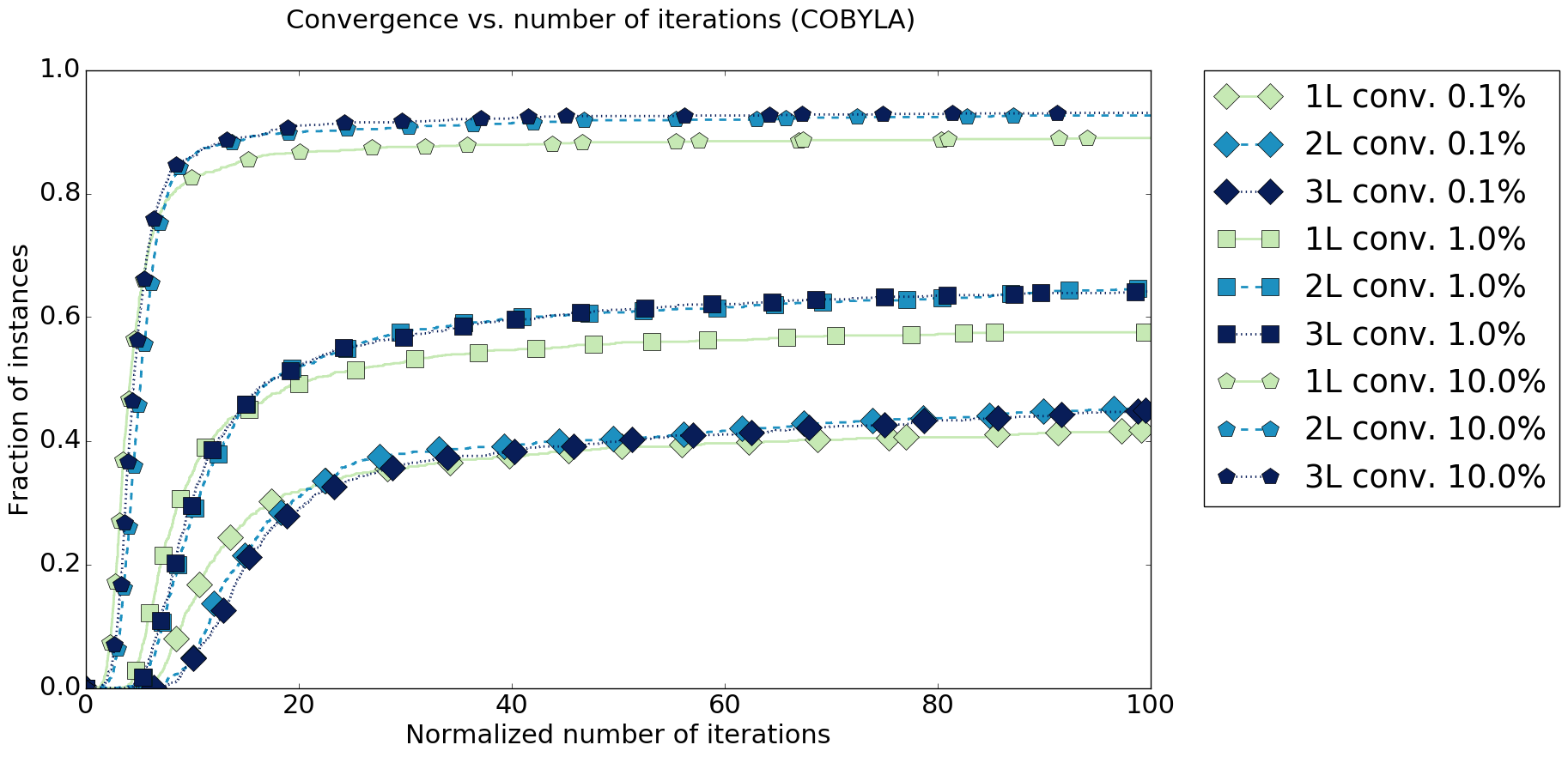}
  }

  \subfloat[RBFOpt.]{
    \includegraphics[width=0.5\textwidth]{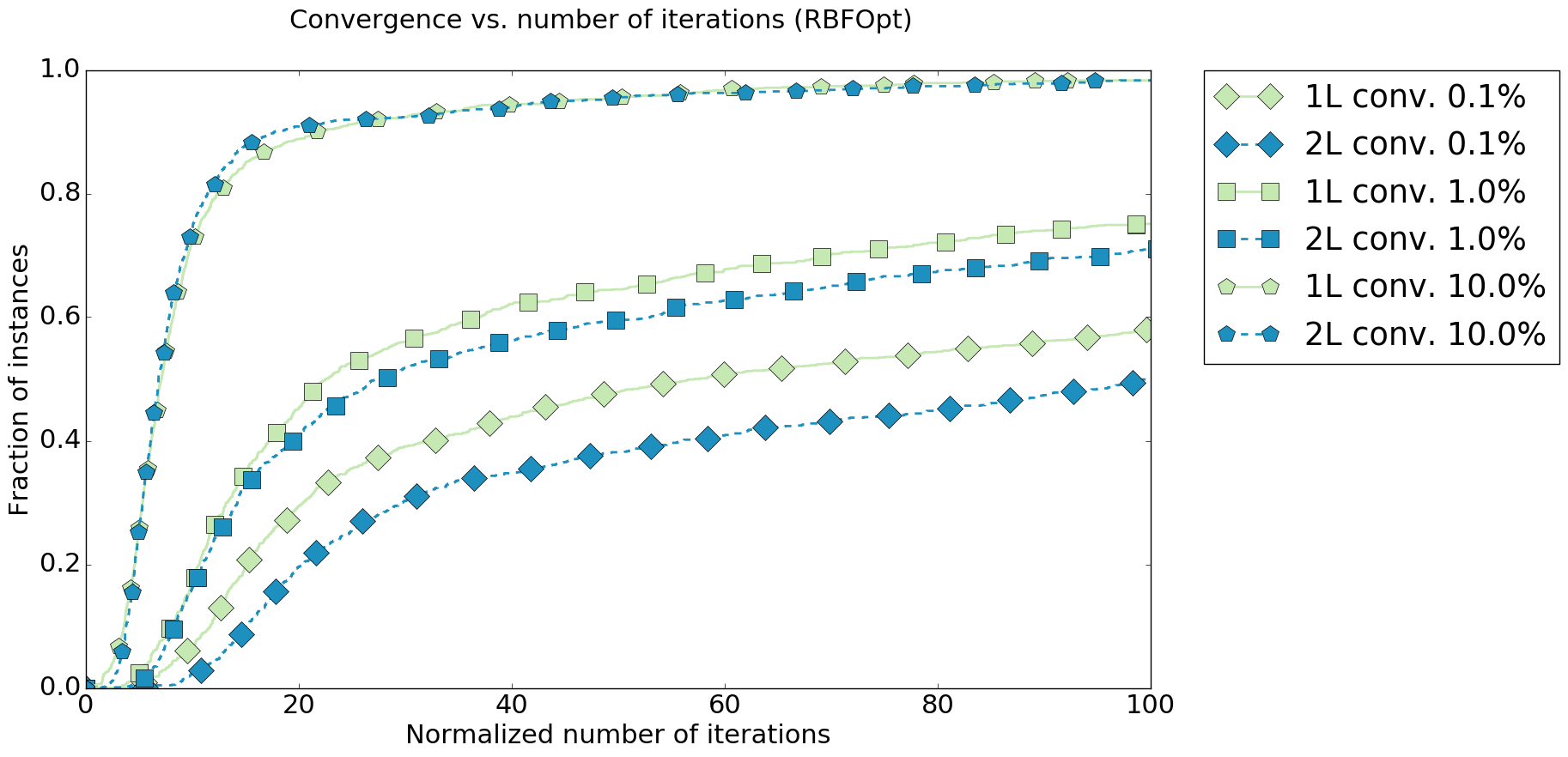}
  }
  
  \caption{Fraction of the instances on which a given algorithm
    converges to the specified tolerance, versus the normalized number
    of iterations.}
  \label{fig:conv_by_iter}
\end{figure}

In Fig.~\ref{fig:conv_by_iter}, we report aggregate results for
COBYLA and RBFOpt over the entire test set. As mentioned earlier,
LBFGS's performance is very close to that of COBYLA. The curves are
drawn for three convergence levels: $\tau \in \{0.001, 0.01,
0.1\}$. The $x$-axis indicates the normalized number of iterations
(i.e., equivalent gradient iterations), the $y$-axis reports the
fraction of instances on which the optimization algorithm converges up
to a specified tolerance.

We can see from Fig.~\ref{fig:conv_by_iter} that the local
optimization algorithm (COBYLA) plateaus after a relatively small
number of normalized iterations, whereas the global optimization
algorithm (RBFOpt) continues improving and in the long run achieves
convergence on more instances. LBFGS shows the same behavior as
COBYLA, and this suggests that the local optimization algorithms are
stuck in a local minimum. However, all algorithms fail to converge to
high accuracy in a large fraction of the instances. Interestingly,
using two layers of the variational form seems to be better when
relying on a local optimization method, but worse when employing the
global algorithm of RBFOpt. A possible explanation is that RBFOpt
works better when the number of parameters to be optimized is small,
and therefore does not benefit from the enlarged search space found in
the case of several layers of the variational form. In other words,
this fact may stem from properties of the optimization algorithm,
rather than the variational form itself. It is therefore not clear
whether the additional layers, which introduce entanglement, truly
help. This will be discussed more in detail in Section
\ref{s:entnoent}.

\begin{figure}[tb]
  \subfloat[Constrained Optimization By Linear Approximation (COBYLA).]{
    \includegraphics[width=0.5\textwidth]{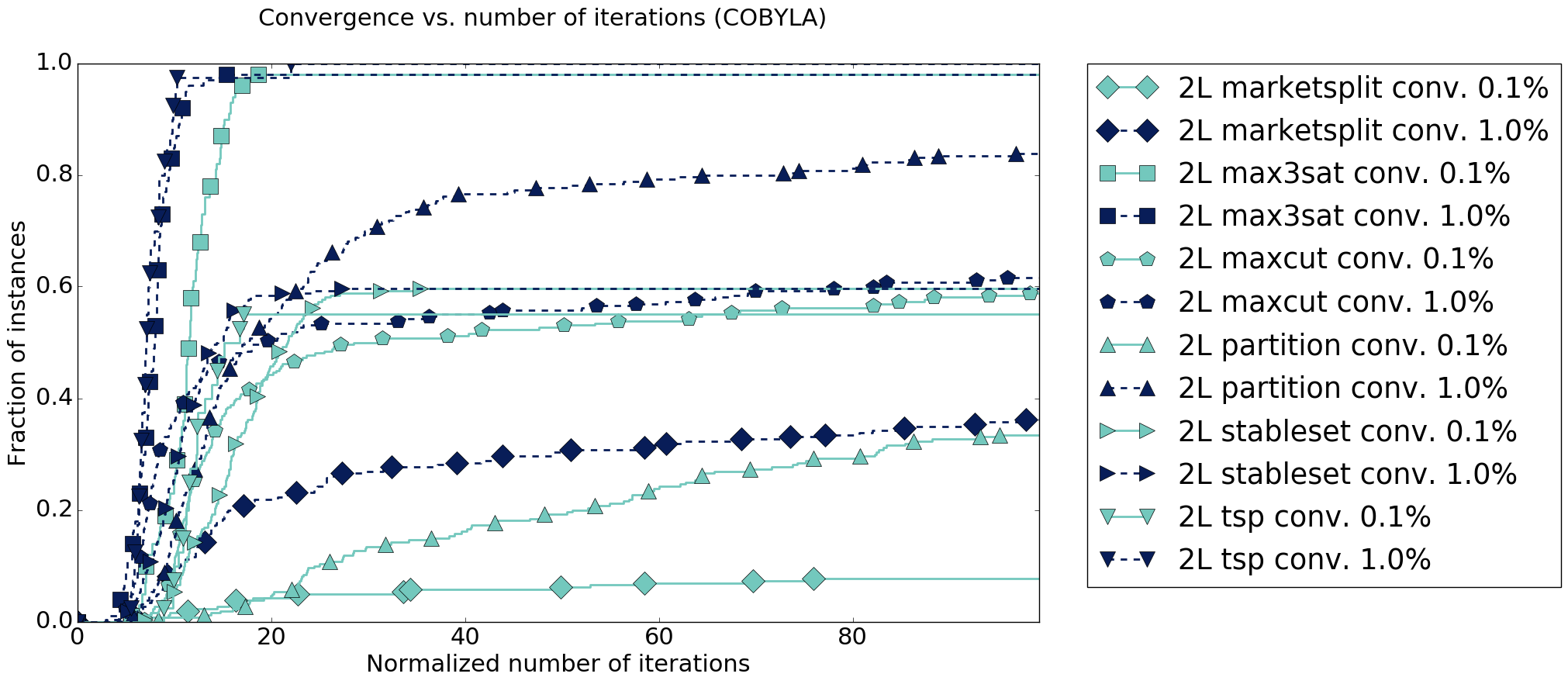}
  }

  \subfloat[RBFOpt.]{
    \includegraphics[width=0.5\textwidth]{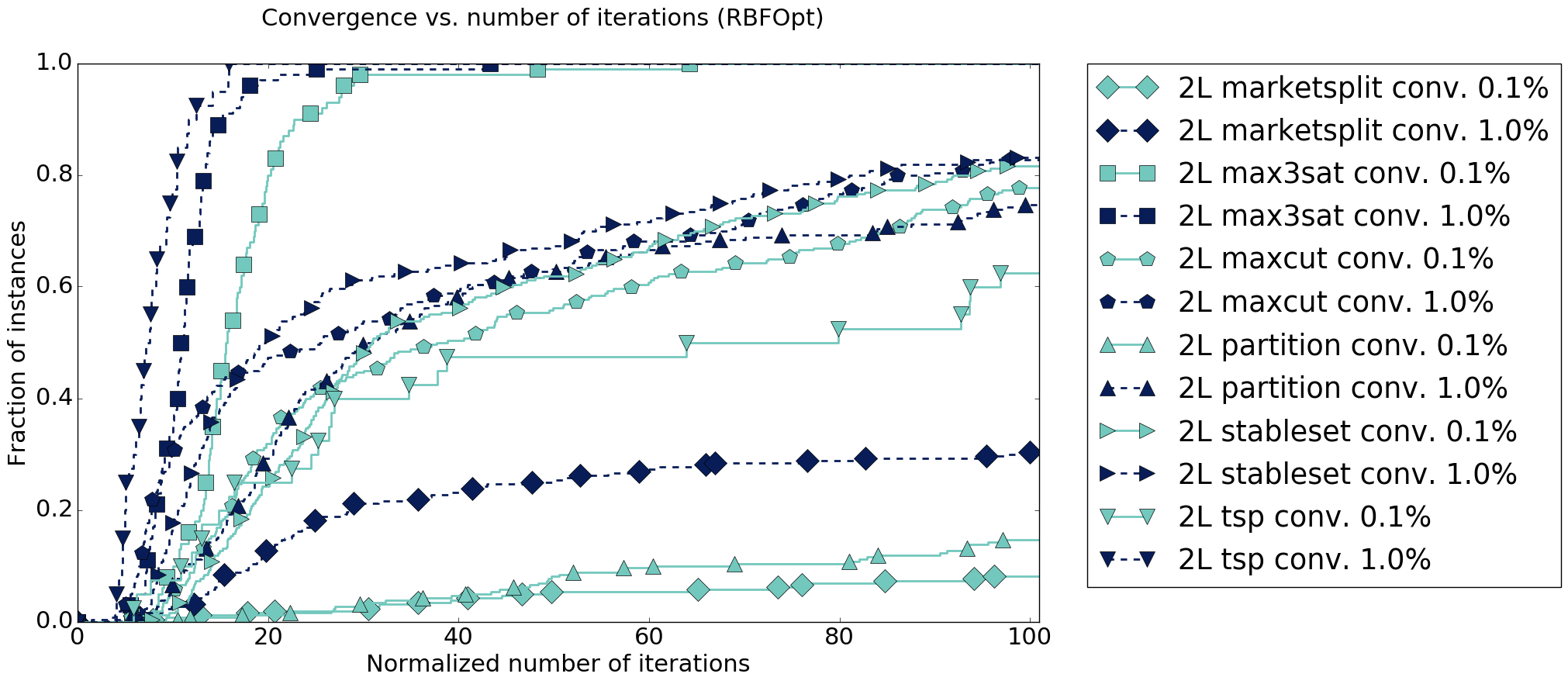}
  }
  
  \caption{Fraction of the instances on which a given algorithm
    converges to the specified tolerance, versus the normalized number
    of iterations. In these plots we employed a variational form with
    two layers.}
  \label{fig:conv_by_iter_prob}
\end{figure}

To understand whether some of this behavior is problem-dependent or
can be observed on the entire set of test instances, we report similar
plots in Fig.~\ref{fig:conv_by_iter_prob}, but now we plot curves
for each instance class. For space reasons, we only provide plots for
two layers of the variational form, but this is representative of the
overall picture. The plots clearly show some problem-dependent
behavior. In particular, some instances are considerably easier than
others: all optimization algorithms excel on {\sc Max3SAT}, but
struggle on {\sc Marketsplit}. The most difficult problem classes are
{\sc Marketsplit}, {\sc Partition} and (to a lesser extent) {\sc
  MaxCut} and {\sc TSP}. This is the same ranking in terms of
difficulty that was obtained when applying a classical
Branch-and-Bound algorithm, which seems to indicate that problems that
are hard for IBM ILOG Cplex are also hard for the VQE heuristic. In
the next subsection we perform further numerical experiments to try to
determine what makes certain problems harder.

To summarize the results presented in this subsection, our experiments
indicate that different problem classes have different difficulty
levels, independent of the optimization algorithm (PCG and SPSA, not
reported here, exhibit similar behavior). This may depend on the
specific procedure adopted to generate random instance, resulting in
harder instances for some classes of problems.  These remarks are
consistent with the literature in classical combinatorial
optimization, where the most successful methods to solve problems take
advantage of problem-specific structure. The agnostic nature of VQE,
coupled with classical derivative-free optimization algorithms,
results in alternating performance with mixed results that seem to
match the behavior of classical Branch-and-Bound.

\subsection{Density and eigenvalues}
Table \ref{tab:terms} shows that the hardest problems have more
distinct eigenvalues than easier problems ({\sc TSP}, that exhibits a
large number of distinct eigenvalues, seems not too difficult in
Fig.~\ref{fig:conv_by_iter_prob}, but the analysis in subsequent
sections indicates that the approximation ratio of the solution found
by VQE is fairly large). These problems also seem to have
Hamiltonians with more terms. It is important to investigate if these
quantities correlate with difficulty.

%% There is another possible explanation for the difficulty of {\sc
%%   Marketsplit} and {\sc Partition} behavior, which is partially
%% supported by data: Fig.~\ref{fig:value_dist} shows that most
%% solutions for instances of these two problems have low function value,
%% and just a handful of binary strings have very large function
%% value. More importantly, the curves for {\sc Marketsplit} and {\sc
%%   Partition} have the smallest slope for a large part of the $x$
%% axis. Our methodology to evaluate convergence considers the
%% improvement with respect to the initial point: since the slope of the
%% curves that plot the distribution of function values is slow, {\sc
%%   Marketsplit} and {\sc Partition} have many solutions that offer
%% little improvement with respect to the initial point, and progress
%% toward the minimum energy state could be slow. However, the curve for
%% {\sc MaxCut} in \ref{fig:value_dist} does not have the same property,
%% and {\sc MaxCut} is the hardest problem after {\sc Marketsplit} and
%% {\sc Partition}, which seems to contradict the explanation in terms of
%% the distribution of function values.

In the classical setting, the density of the quadratic objective
function matrix is an important factor in determining the difficulty
of an instance. When discussing quantum Hamiltonians, the most natural
proxy is the number of Pauli terms that appear in the summation
defining the Hamiltonian. Because we are using a nonlinear
optimization tool to determine the ground state, it is also
conceivable that the number of distinct eigenvalues of the Hamiltonian
could be a good indicator of the difficulty of \eqref{eq:vqe} (since
every eigenvector is a stationary point of the optimization
problem). To test these conjectures, we run some experiments on
Hamiltonians consisting of a weighted summation of random pairs of
Pauli $Z$. We fix the number of pairs in each Hamiltonian o
$10,20,\dots,100$, and the weights are chosen uniformly at random in
$\{-1, 1\}$ in the first set of experiments, $[-1, 1]$ in the second
set of experiments (notice that the first is a discrete set, the
second is a real interval). The number of qubits varies between $10$
and $18$, and for each combination of parameters we generate 20 random
Hamiltonians. The average number of unique eigenvalues of these
Hamiltonians is given in Table \ref{tab:unique_eig_randz} (for space
reasons, we report only for $q$ even). As expected, the Hamiltonians
in the second set of experiments (random weights in $[-1, 1]$) have
many more unique eigenvalues than in the first set of experiments,
even if the number of Pauli terms is the same. Indeed, for large
enough number of Pauli terms, the Hamiltonians with random weights in
$[-1, 1]$ have the maximum number of distinct eigenvalues ($2^{q-1}$,
since by construction for every eigenvalue $\lambda$, $-\lambda$ is
also an eigenvalue).

\begin{table}[tbh]
  \centering
  {\scriptsize
  \begin{tabular}{|l|c|c|c|c|c|c|c|c|c|c|}
    \hline 
    \# & \multicolumn{10}{c|}{\# of qubits $q$} \\
    \cline{2-11}
    Pauli & \multicolumn{5}{c|}{weights in $\{-1,1\}$} & \multicolumn{5}{c|}{weights in $[-1,1]$} \\
    \cline{2-11}
    terms & 10 & 12 & 14 & 16 & 18 & 10 & 12 & 14 & 16 & 18 \\
    \hline
    10 & 9 & 9 & 10 & 10 & 11 & 336 & 390 & 633 & 755 & 896 \\
    20 & 13 & 14 & 16 & 17 & 18 & 499 & 1843 & 6246 & 19251 & 45056 \\
    30 & 16 & 18 & 20 & 20 & 22 & 512 & 1997 & 7680 & 30310 & 95027 \\
    40 & 19 & 21 & 22 & 24 & 26 & 512 & 2048 & 8192 & 31948 & 117964 \\
    50 & 21 & 23 & 25 & 27 & 29 & 512 & 2048 & 8192 & 32768 & 131072 \\
    60 & 23 & 26 & 28 & 30 & 33 & 512 & 2048 & 8192 & 32768 & 131072 \\
    70 & 24 & 28 & 31 & 32 & 35 & 512 & 2048 & 8192 & 32768 & 131072 \\
    80 & 26 & 30 & 32 & 34 & 37 & 512 & 2048 & 8192 & 32768 & 131072 \\
    90 & 14 & 32 & 34 & 36 & 39 & 512 & 2048 & 8192 & 32768 & 131072 \\
    100 & 14 & 33 & 35 & 38 & 41 & 512 & 2048 & 8192 & 32768 & 131072 \\
    \hline
  \end{tabular}
  }
  \caption{Average number (rounded to the nearest integer) of unique
    eigenvalues in the Hamiltonian depending on the number of Pauli
    terms in the summation defining the Hamiltonian.}
  \label{tab:unique_eig_randz}
\end{table}

\begin{figure}[tb]
  \subfloat[Hamiltonians with weights in $\{-1,-1\}$.]{
    \includegraphics[width=0.5\textwidth]{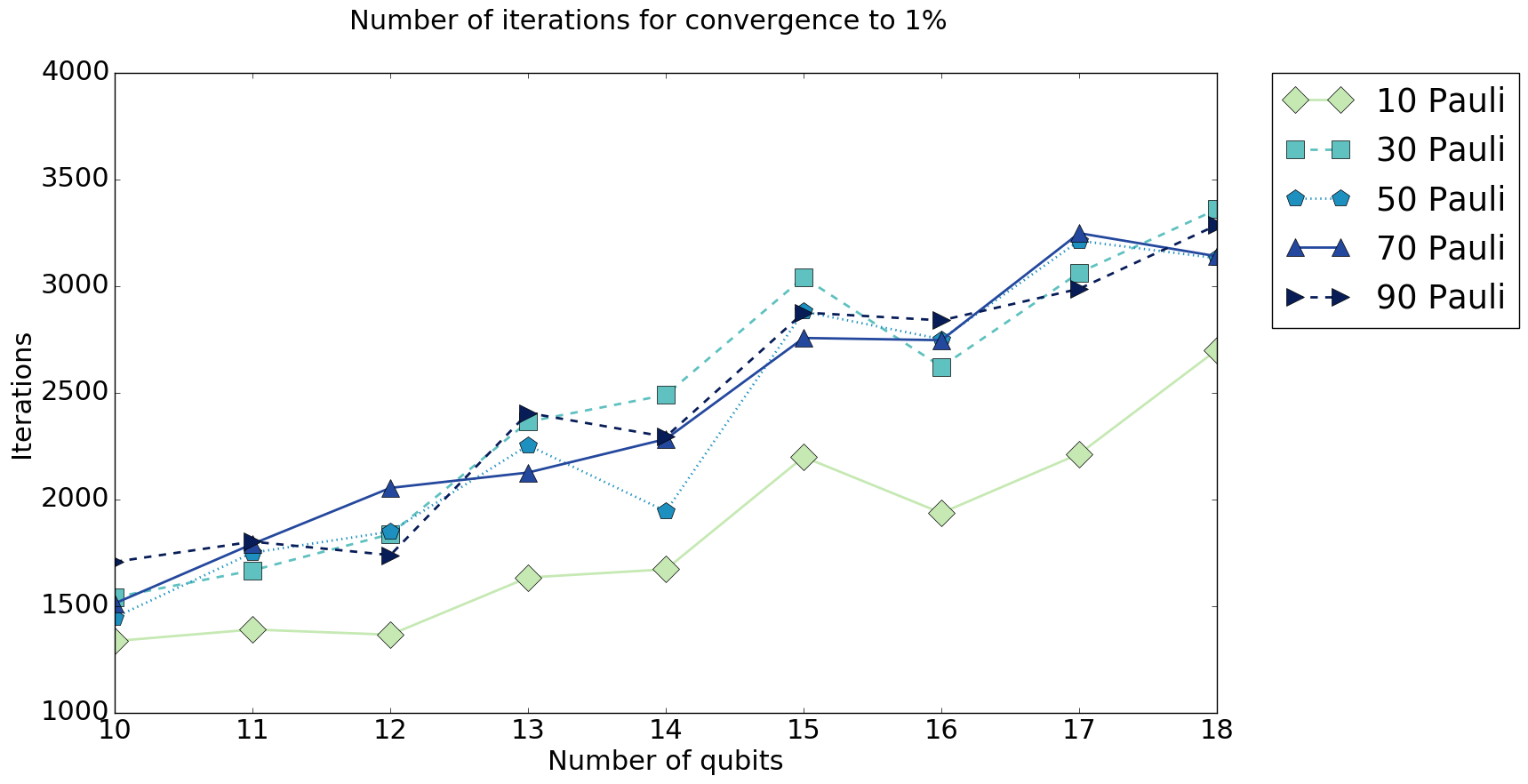}
  }

  \subfloat[Hamiltonians with weights in {$[-1,1]$}.]{
    \includegraphics[width=0.5\textwidth]{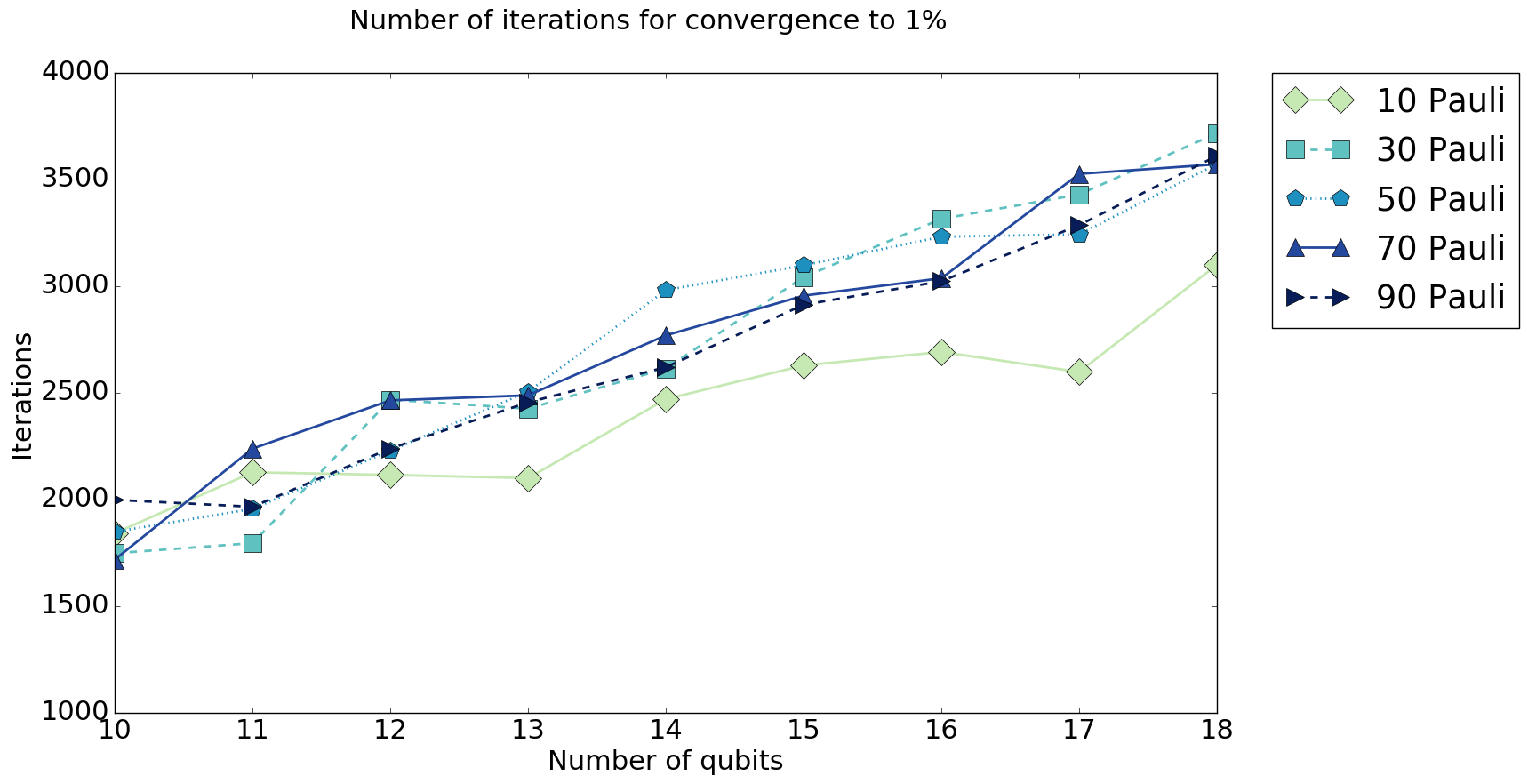}
  }
  
  \caption{Average number of iterations for convergence to $\tau =
    0.01$ for Hamiltonians with a different number of Pauli
    terms, using COBYLA and a variational form with $2$ layers.}
  \label{fig:randz_cobyla}
\end{figure}

Running VQE on these Hamiltonians shows that density is not a good
indicator of difficulty, but the number of eigenvalues is. In
Fig.~\ref{fig:randz_cobyla} we report convergence with $\tau = 0.01$
and a variational form with 2 layers using COBYLA; results with other
optimizers or number of layers are similar. The graph suggests that
problems with very small number of Pauli terms (e.g., 10) are easy
across all sizes but as soon as the number increases it is no longer
possible to detect a strong correlation between number of terms and
difficulty. However, the number of distinct eigenvalues affects
difficulty (remember from Table \ref{tab:unique_eig_randz} that there
is saturation of the number of eigenvalues for more than $\approx 30$
Pauli terms in the Hamiltonian with weights in $[-1,1]$, hence we
cannot expect problems to get more difficult when they have more than
$30$ terms).

To support this conclusion, we compare the number of iterations for
convergence for increasing number of Pauli terms using a
nonparametric statistical test known as the Friedman test. The groups
(algorithms) compared correspond to the number of Pauli terms in the
Hamiltonian, for uniform weights in $\{-1,1\}$ and in $[-1,1]$. We use
confidence $\alpha = 0.95$; the null-hypothesis (no differences
between the variables) is rejected, p-value $1.11e^{-16}$, and we
perform pairwise comparisons in the post-hoc analysis. The post-hoc
analysis clearly indicates that for the same number of terms in the
Hamiltonian, problems with uniform weights in $[-1,1]$, which have
more distinct eigenvalues, take longer to converge. We report a subset
of the results in Table \ref{tab:posthoc}.

{\renewcommand{\arraystretch}{1.2}%
\begin{table}[htb]
  \begin{tabular}{|c|c|l|*{10}{c|}}
    \hline
    \multicolumn{3}{|c|}{} & \multicolumn{10}{c|}{Number of Pauli terms} \\
    \cline{4-13}
\multicolumn{3}{|c|}{} & \multicolumn{5}{c|}{Weights $\{-1,1\}$} & \multicolumn{5}{c|}{Weights $[-1,1]$} \\
\cline{4-13}
\multicolumn{3}{|c|}{} & 10& 30&  50&  70&  90& 10&  30&  50&  70&  90 \\
\hline
\parbox[t]{3mm}{\multirow{10}{*}{\rotatebox[origin=c]{90}{Number of Pauli terms}}}
& \parbox[t]{3mm}{\multirow{5}{*}{\rotatebox[origin=c]{90}{Weights $\{-1,1\}$}}} & 10 &   & - &  - &  - &  - &  -&   -&   -&   -&   - \\
& & 30 & + &   &  = &  = &  = &  =&   -&   -&   -&   - \\
& & 50 & + & = &    &  = &  = &  =&   -&   -&   -&   - \\
& & 70 & + & = &  = &    &  = &  =&   -&   -&   -&   - \\
& & 90 & + & = &  = &  = &    &  =&   -&   -&   -&   - \\
\cline{2-13}
& \parbox[t]{3mm}{\multirow{5}{*}{\rotatebox[origin=c]{90}{Weights $[-1,1]$}}} & 10 & + & = &  = &  = &  = &   &   -&   -&   -&   - \\
& & 30 & + & + &  + &  + &  + &  +&    &   =&   =&   = \\
& & 50 & + & + &  + &  + &  + &  +&   =&    &   =&   = \\
& & 70 & + & + &  + &  + &  + &  +&   =&   =&    &   = \\
& & 90 & + & + &  + &  + &  + &  +&   =&   =&   =&     \\
\hline
  \end{tabular}
  \caption{Pairwise comparison of the number of iterations for
    convergence. A ``$+$'' in row $i$ and column $j$
    indicates that in experiment $i$ the Friedman test detected more
    iterations than in experiment $j$; vice versa with a ``$-$''; no
    difference is detected with a ``$=$''. The two-digit numbers
    labeling column and rows indicate the number of Pauli terms in the
    experiment.}
  \label{tab:posthoc}
\end{table}
}

To conclude, our experiments with random Hamiltonians obtained as sum
of pairs of $Z$s indicate that we can expect the difficulty of a
problem instance to increase with the number of distinct
eigenvalues. This, in turn, can be related to the number of Pauli
terms in the summation defining the Hamiltonian, depending on the
distribution of the weights. The results are consistent with our
observations in the preceding sections. From a practical standpoint,
unfortunately computing the number of distinct eigenvalues is not an
easy task unless the Hamiltonian is known in the full Hilbert space.

\subsection{Impact of problem size}
\begin{figure}[tb]
  \subfloat[Constrained Optimization By Linear Approximation (COBYLA).]{
    \includegraphics[width=0.5\textwidth]{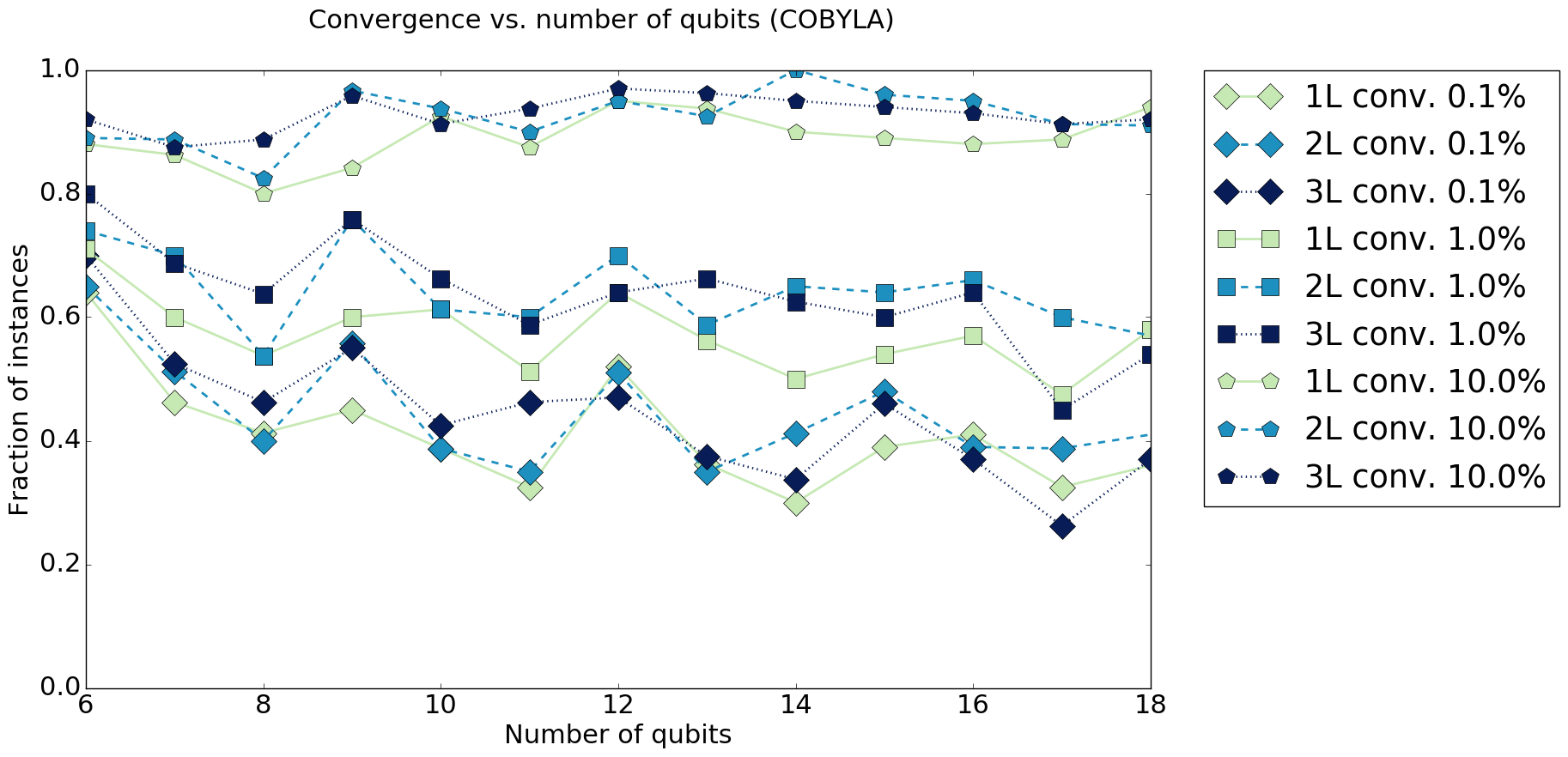}
  }

  \subfloat[RBFOpt.]{
  \includegraphics[width=0.5\textwidth]{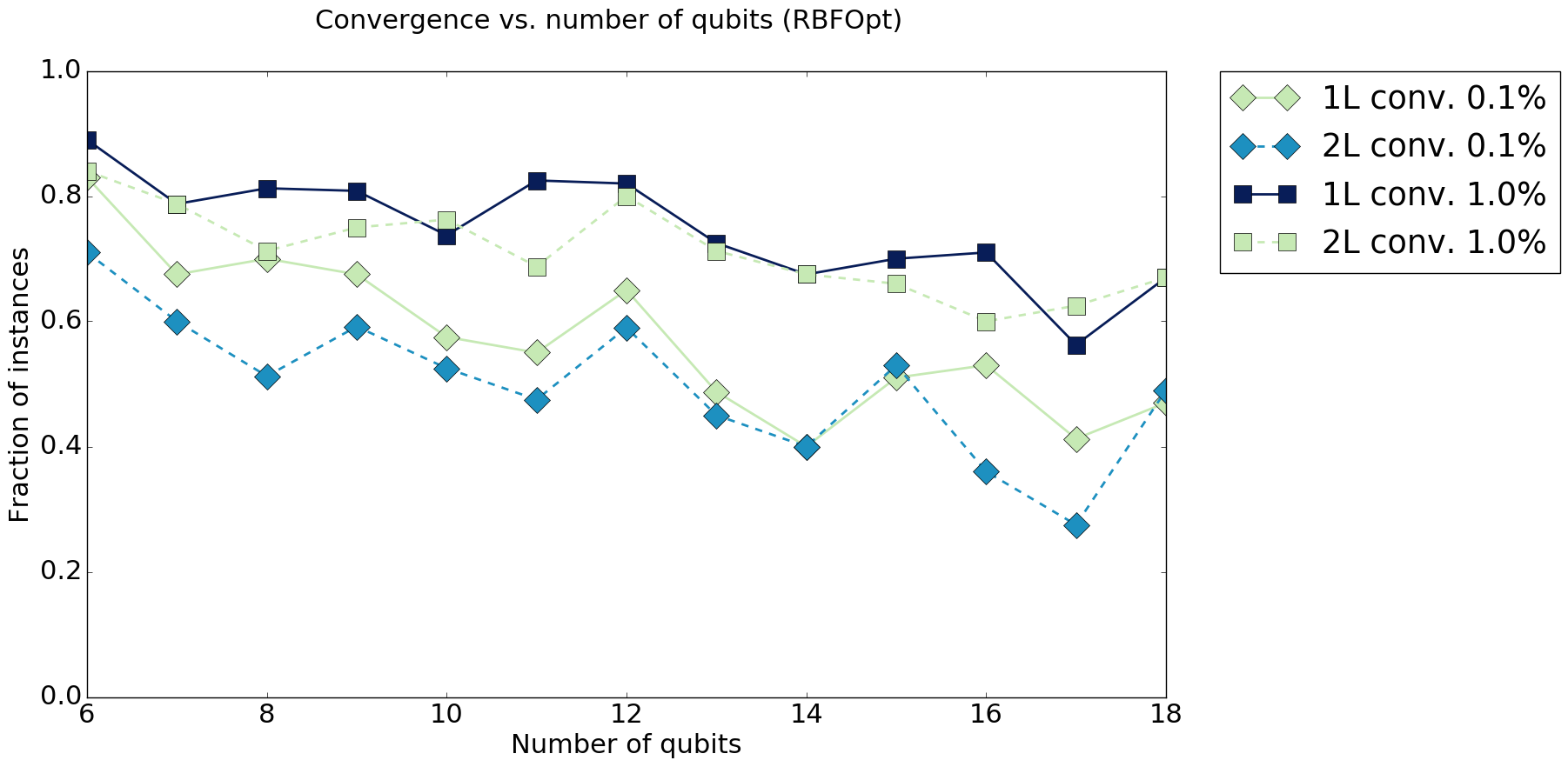}
  }
  
  \caption{Fraction of the instances on which a given algorithm
    converges to the specified tolerance, versus the number of
    qubits.}
  \label{fig:conv_by_size}
\end{figure}

We now study the impact of problem size, i.e., number of qubits, on
the performance of VQE. Fig.~\ref{fig:conv_by_size} reports the
fraction of problem istances on which convergence to a specified level
of tolerance is attained, with problem size varying from 6 to 18
qubits. Results are aggregated over all problem classes.

The plot indicates a slight downward trend in the convergence rate as
problem size increases. We remark that in these experiments the
scaling of the number of iterations that each algorithm is allowed to
run is the same as in the previous section, namely $100(q\ell+1)$
iterations where $q$ is the number of qubits. Hence, the plot suggests
that increasing the number of iterations linearly in the number of
qubits is not sufficient to maintain constant the fraction of
instances on which convergence to high accuracy is attained. Indeed,
as remarked in the previous section, Fig.~\ref{fig:conv_by_iter}
shows diminishing returns for increased iteration number, and some
algorithims reach a plateau after a certain number of iterations;
Fig.~\ref{fig:conv_by_size} provides the additional information that
problem instances become harder to solve as the qubit count
increases. The data gathered from our experiments is not sufficient to
extract an overall trend and determine what is the correct scaling of
the number of iterations to maintain a constant fraction of instances
on which converge is achieved.

\begin{figure}[tb]
  \subfloat[Constrained Optimization By Linear Approximation (COBYLA).]{
    \includegraphics[width=0.5\textwidth]{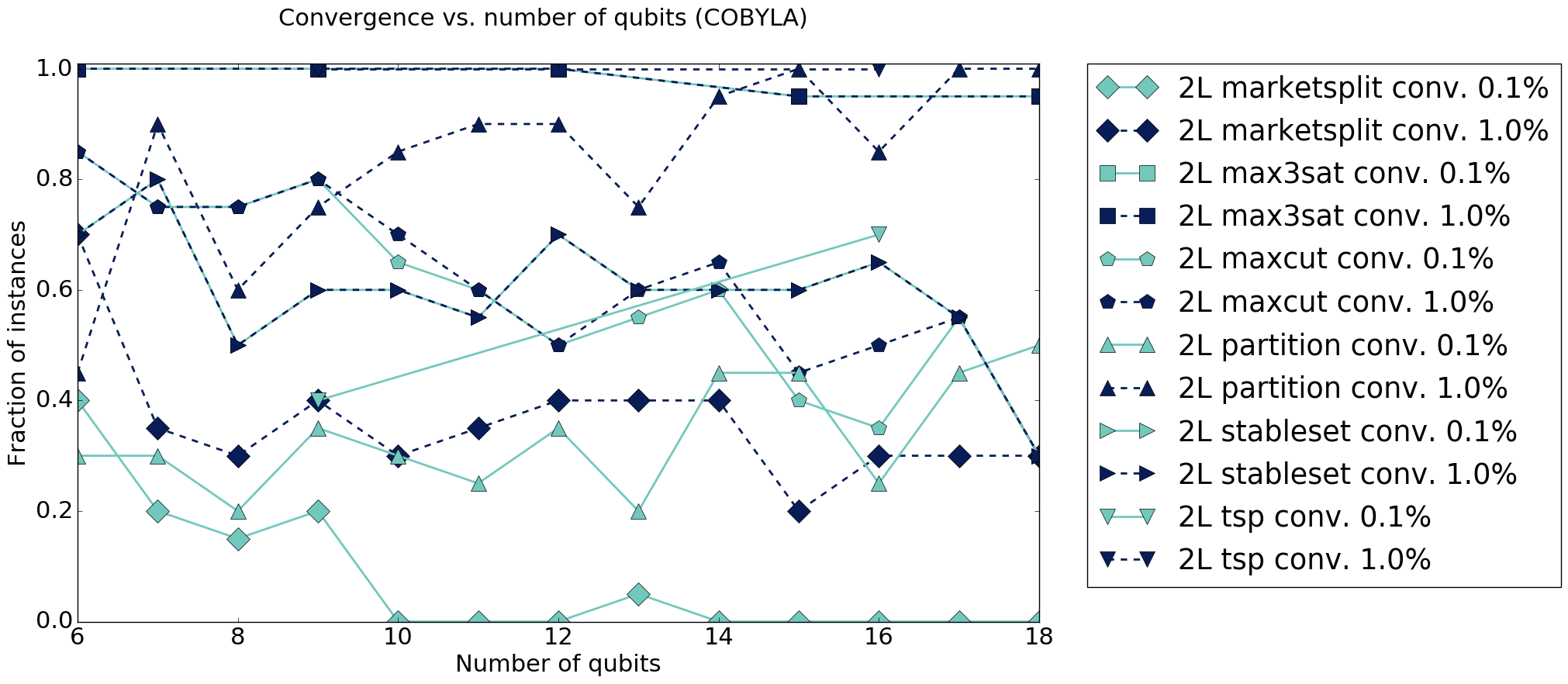}
  }

  \subfloat[RBFOpt.]{
    \includegraphics[width=0.5\textwidth]{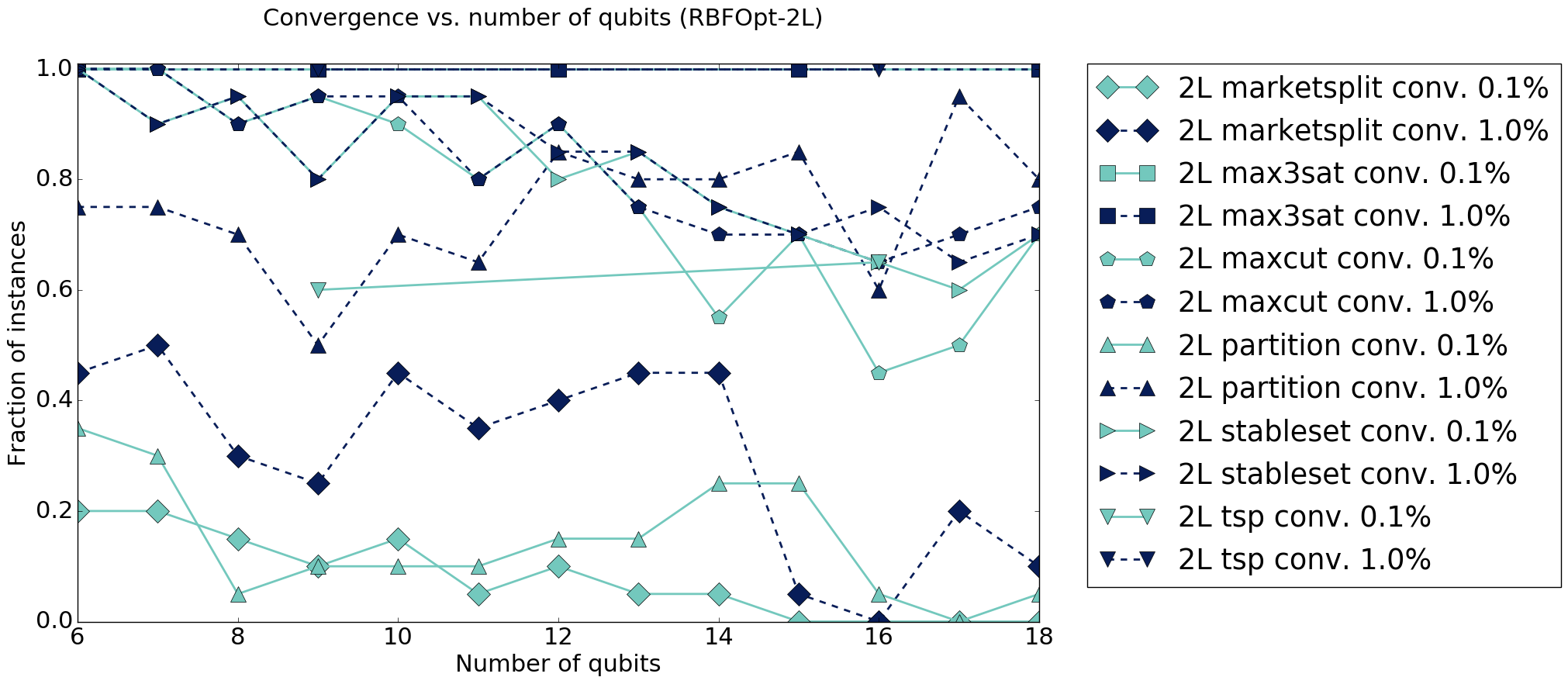}
  }
  
  \caption{Fraction of the instances on which a given algorithm
    converges to the specified tolerance, versus the number of
    qubits. In these plots we employed a variational form with two
    layers.}
  \label{fig:conv_by_size_prob}
\end{figure}

Similar plots that provide a curve for each problem class
(Fig.~\ref{fig:conv_by_size_prob}) indicate that the downward trend is
not observed among all problem classes: specifically, the performance
of all optimization algorithms on Partition and TSP does not seem to
deteriorate for increasing problem size. However, this is the case for
the remaining problems, and this leads to our observations for the
average case.

\subsection{Approximation ratio}
\begin{figure}[tb!]
  \subfloat[{\sc Partition}.]{
    \includegraphics[width=0.5\textwidth]{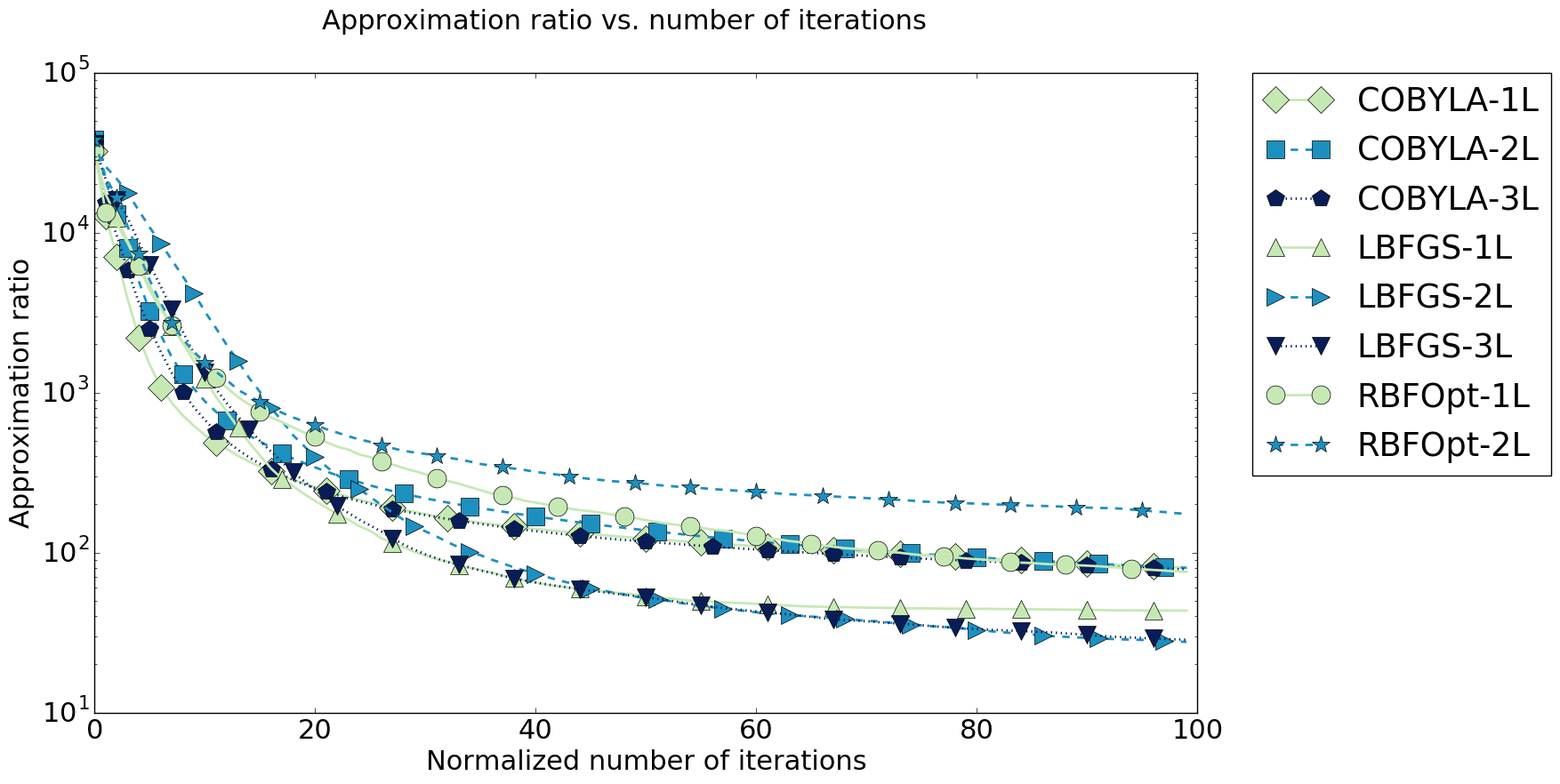}
  }

  \subfloat[{\sc MaxCut}]{
    \includegraphics[width=0.5\textwidth]{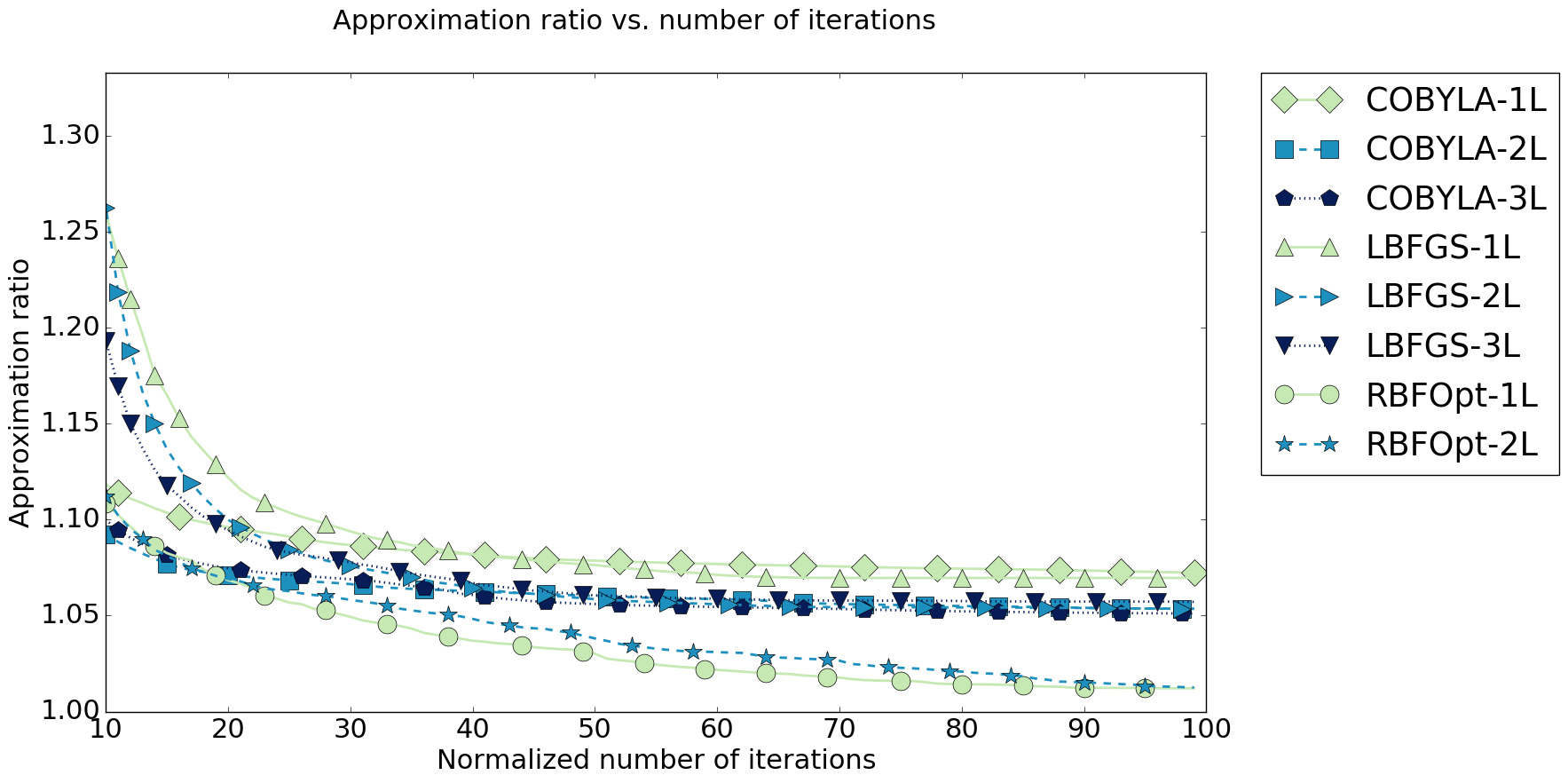}
  }

  \caption{Average approximation ratio (geometric average) versus the
    normalized number of iterations. The $y$ axis in the graph for
    {\sc Partition} is on log scale.}
  \label{fig:approx_ratio}
\end{figure}

The convergence profiles shown in previous sections have the useful
properties of being normalized with respect to the initial point and
invariant to constant shifts to the diagonal of the Hamiltonian. A
more commonly used metric to assess the performance of optimization
methods (especially from a theoretical point of view) is the
approximation ratio, defined as the value $(1 + \epsilon)$ such that
the algorithm attains a solution of value at most $f(x^\ast)(1 +
\epsilon)$, where $f(x^\ast)$ is the optimum value.  We report the
evolution of the approximation ratio for {\sc {Partition}} and {\sc
  MaxCut} in Fig.~\ref{fig:approx_ratio}. We remark that some of the
test problems have energy values unrestricted in sign, i.e., the
classical optimization algorithm explores points with positive and
negative value. The approximation ratio only makes sense for
nonnegative values. Notice that the minimum eigenvalue of the
Hamiltonian can be negative, in particular for all $\max$ optimization
problems converted to a $\min$ problem by taking the negative of the
objective function. Therefore, in our graphs we only report iterations
for which the energy value of the quantum state has the same sign as
the minimum eigenvalue of the Hamiltonian, and for problems with
negative minimum eigenvalue we plot the ratio $f(x^\ast)/f(x)$ rather
than $f(x)/f(x^\ast)$; this way, all graphs are decreasing and lower
bounded by $1$. The average across all instances is taken using the
geometric average, which is more suitable in this context since the
approximation ratio is a multiplicative quantity rather than additive.

On {\sc Partition} the approximation ratio is very large, consistent
with the observation that this problem class appears to be
difficult. The performance on {\sc Marketsplit} is very similar
(additional graphs are given in the Appendix). On {\sc MaxCut}, the
global optimization algorithm eventually reaches an approximation
ratio close to $1$, but convergence is slow; the local optimization
algorithms are stuck in local optima with ratio $\ge 1.05$. {\sc
  StableSet} is similar to {\sc MaxCut}, whereas on {\sc Max3SAT} all
algorithms quickly attain ratio very close to 1, consistent with our
previous observations. {\sc TSP} starts with approximation ratios
$\approx 10^3$ but quickly reaches values $\le 10$, even though no
algorithm attains $1$. Since the original formulation for {\sc TSP} is
heavily constrained (transformed to unconstrained by penalizing
constraint violations), the initial point given to the classical
optimization algorithm is likely infeasible (i.e., it does not satisfy
the constraints) and incurs heavy penalties that affect its energy
value; the optimization then moves towards feasibility, which explains
the large initial approximation ratios that quickly decrease as the
feasible region is reached.

\subsection{Probability of sampling the optimal solution}
\label{s:sampling}
So far, we have been concerned with studying the speed with which
optimization algorithm find a quantum state with an optimal or
close-to-optimal energy value, as evaluated according to a given
Hamiltonian. We remark that the energy of a quantum state corresponds
to the expected objective function value of the binary solutions that
can can be sampled from that quantum state. This can be easily
verified: let $H$ be Hamiltonian encoding of a combinatorial problem
with objective function $f : \{0,1\}^q \to \R$. Then if $\ket{\psi}$
is a basis state, it corresponds to a binary string $z$, and we must
have $\bra{\psi} H \ket{\psi} = H_{z,z} = f(z)$, where by $H_{z,z}$ we
denote the element of $H$ whose row and column are indexed by
$z$. Furthermore, recall that Hamiltonians for combinatorial problems
encoded as binary problems use only Pauli $Z$ operators, resulting in
a diagonal Hamiltonian. Therefore, for a general state $\ket{\psi}$,
we can write:
\begin{align*}
  \bra{\psi} H \ket{\psi} &= \bra{\psi} \text{diag}(H) \ket{\psi} = \sum_{z \in \{0,1\}^q} \braket{\psi_z}{\psi_z} H_{z,z} \\
  &= \sum_{z \in \{0,1\}^q} \Pr(\ket{\psi} = z) H_{z,z} \\
  &= \sum_{z \in \{0,1\}^q} \Pr(\ket{\psi} = z) f(z).
\end{align*}
In the above expression, $\Pr(\ket{\psi} = z)$ is the probability of
sampling $z$ when performing a measurement of all the $q$ qubits from
state $\ket{\psi}$. 

Because of this relationship, it is in principle possible (although
unlikely) that VQE produces quantum states that have a high
probability of sampling an optimal binary string $z$, while the energy
of the quantum state is larger than the optimum value; see
\cite{farhi2014quantum} for a discussion of concentration of the
probability distribution. We analyze this possibility in our next set
of experiments. More specifically, we look at how the probability of
sampling an optimal solution increases as the optimization algorithm
progresses. In the spirit of the plots reported in previous section,
given a convergence level $\rho$, we compute the fraction of instances
in which a quantum state that has probability at least $\rho$ of
sampling the optimal binary string has been observed within a certain
number of normalized iterations. We remark that this yields
``optimistic'' graphs, because it yields nondecreasing curves: in
practice it is possible that the optimization algorithm explores a
quantum state with high probability of sampling the optimal string,
but the algorithm does not stop there and in subsequent iterations
such probability decreases.

\begin{figure}[tb!]
  \subfloat[Constrained Optimization By Linear Approximation (COBYLA).]{
    \includegraphics[width=0.5\textwidth]{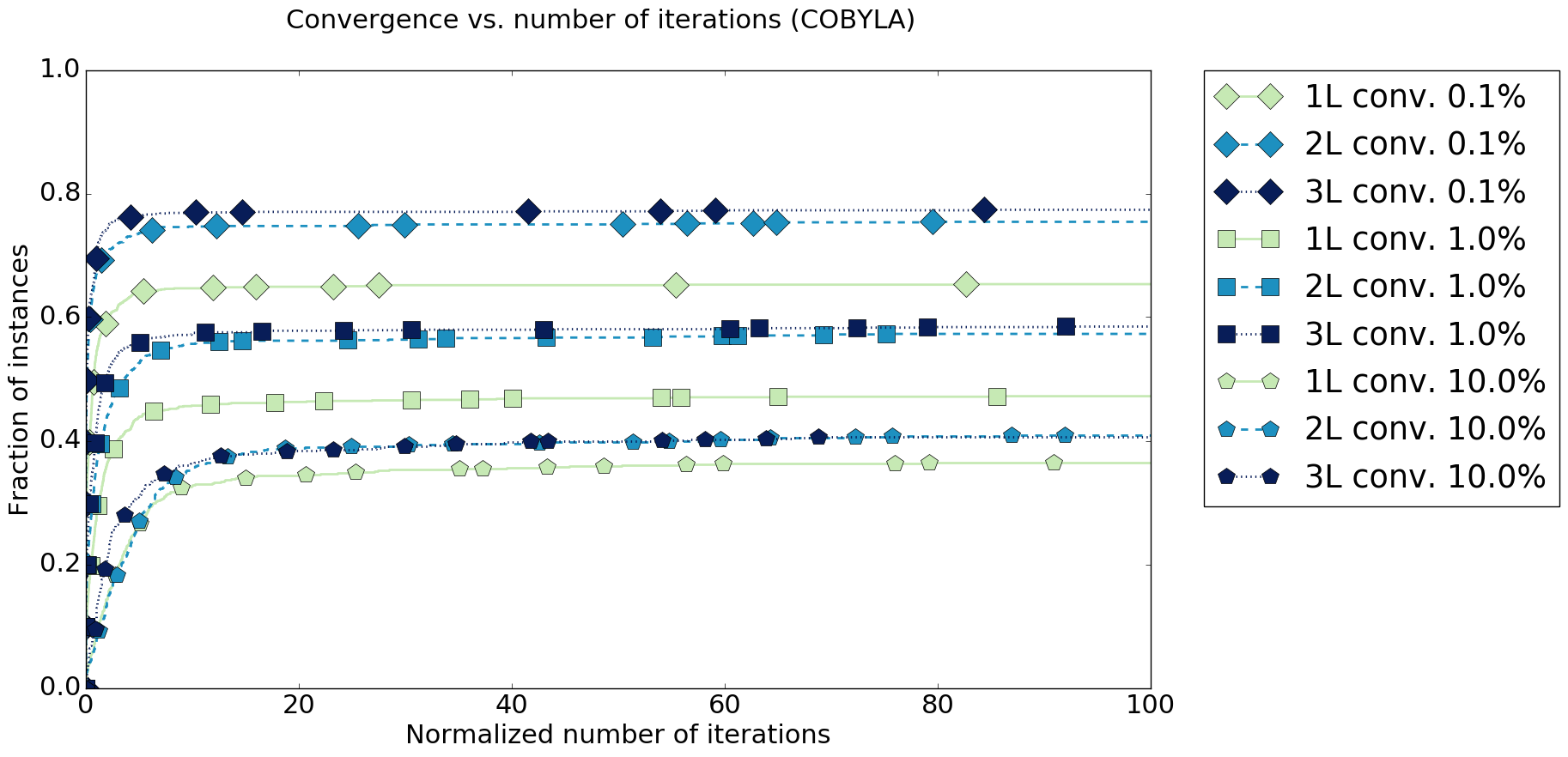}
  }

  \subfloat[RBFOpt.]{
    \includegraphics[width=0.5\textwidth]{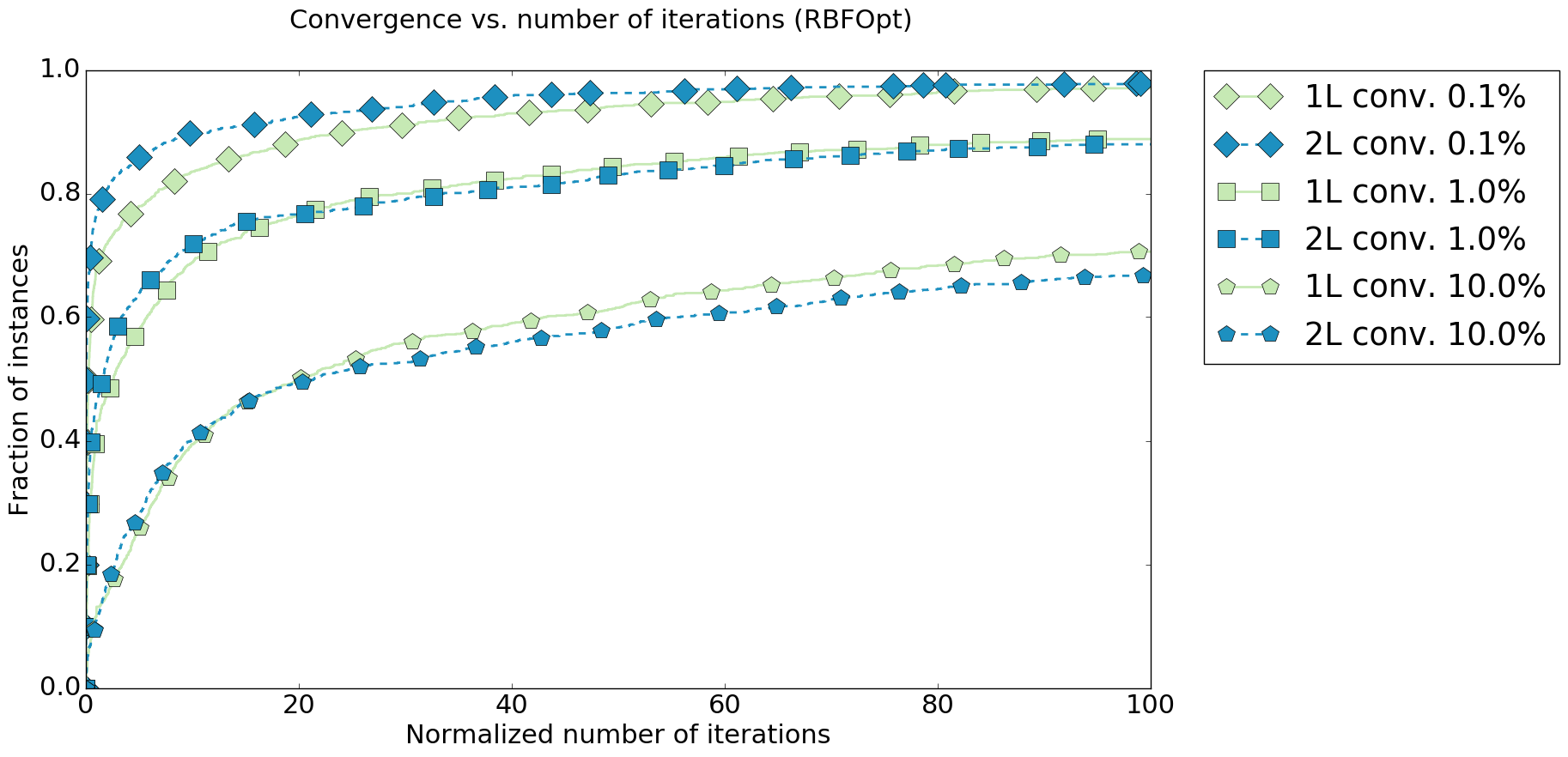}
  }
  
  \caption{Fraction of the instances on which a given algorithm
    explores at least one quantum state with probability of sampling
    the optimal solution greater than a given threshold, versus the
    normalized number of iterations.}
  \label{fig:sample_by_iter}
\end{figure}

\begin{figure}[tb!]
  \subfloat[Constrained Optimization By Linear Approximation (COBYLA).]{
    \includegraphics[width=0.5\textwidth]{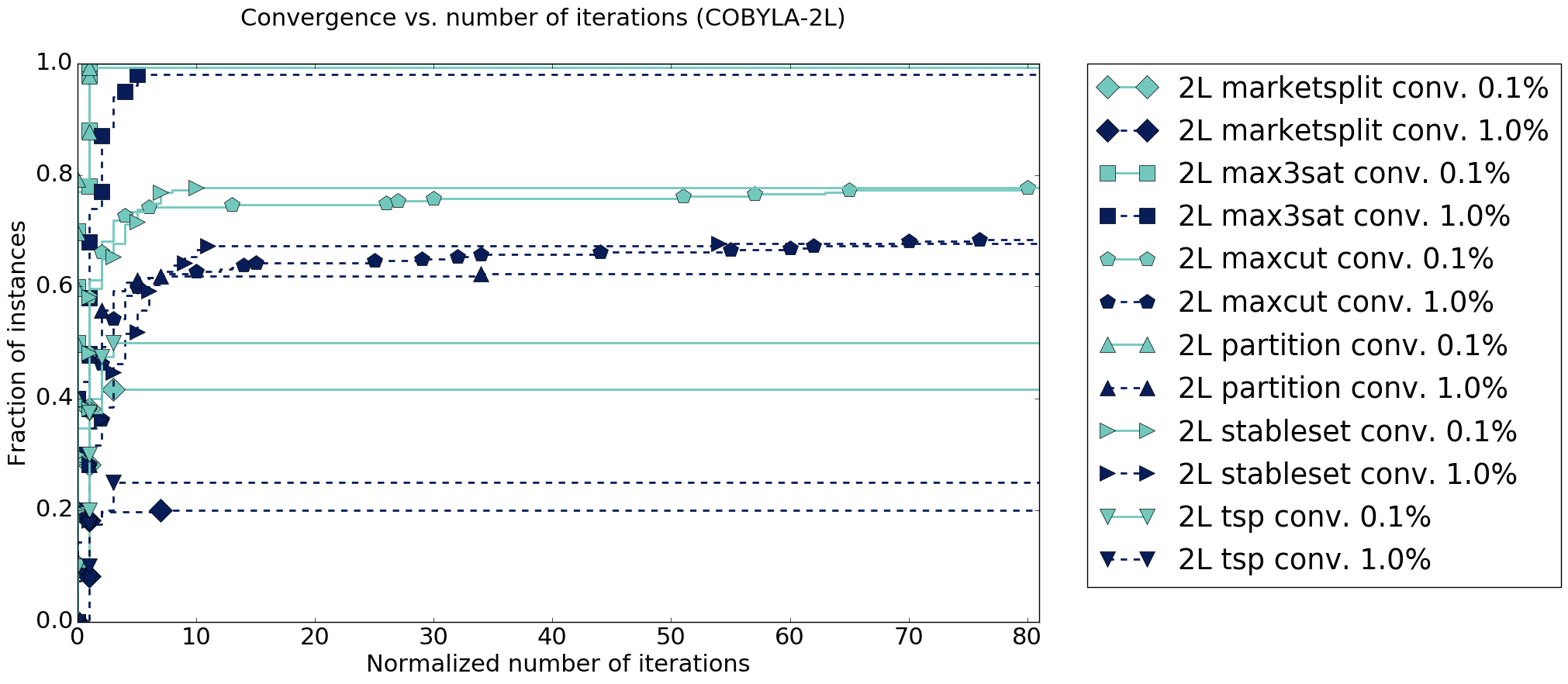}
  }

  \subfloat[RBFOpt.]{
    \includegraphics[width=0.5\textwidth]{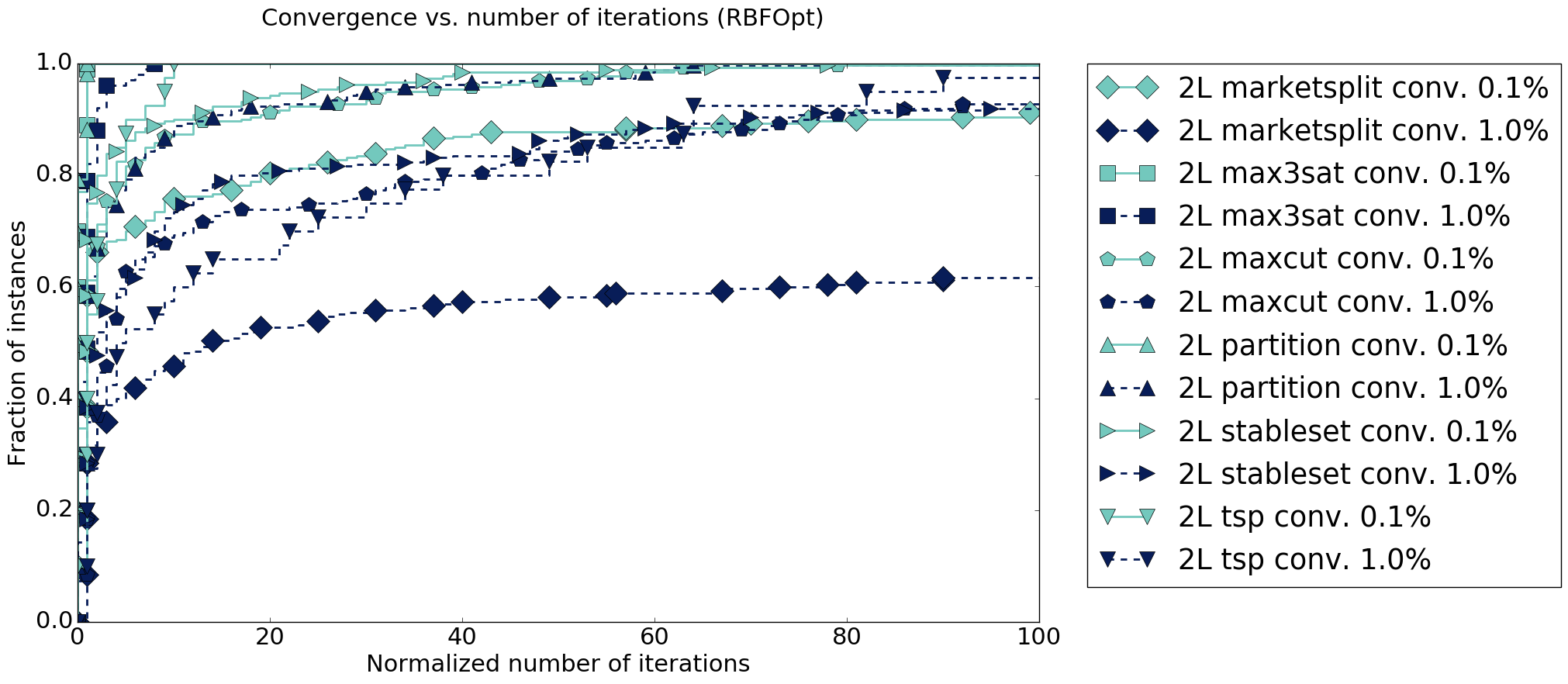}
  }
  
  \caption{Fraction of the instances on which a given algorithm
    explores at least one quantum state with probability of sampling
    the optimal solution greater than a given threshold, versus the
    normalized number of iterations. These plots are generated with a
    variational form with two layers.}
  \label{fig:sample_by_iter_prob}
\end{figure}

Plots generated as discussed above are reported in
Fig.~\ref{fig:sample_by_iter} and \ref{fig:sample_by_iter_prob}. It is
evident from the plots in Fig.~\ref{fig:sample_by_iter} that for the
local optimization algorithm (COBYLA in this graph, but LBFGS is
similar), the fraction of instances in which the probability of
sampling the optimal solution is greater than a given constant
plateaus very quickly. Further iterations of the optimization
algorithm may increase such probability (typically, the energy value
goes down, as observed in previous sections), but the plots show that
this is an improvement on instances for which the optimization
algorithm has already found ``good'' quantum states. The situation is
different for the global algorithm RBFOpt: we observe a steady
increase in the curves, implying that the algorithm explores quantum
states with sufficiently large probability of sampling the optimal
string on more and more instances. However, we remark that for a
global algorithm the ``optimistic'' way of generating these graphs may
have a significant impact: indeed, a global optimization algorithm
will often explore quantum states from unknown regions of the search
space, and for this reason it is more likely to encounter quantum
states that satisfy the convergence criterion, but it may not be able
to detect when one of these states has been found. In other words,
even though RBFOpt explores better quantum states as the iteration
count increases, it may not be able to indicate from which quantum
state the optimal string can be sampled. In any case, these plots
indicate that the local optimization algorithms can quickly get stuck
in local minima that do not contain quantum states likely to yield the
optimal binary string. This is not suprising: due to their hardness,
we expect NP-hard combinatorial optimization problem to give rise to
Hamiltonians associated with highly nonconvex energy landscapes.

Fig.~\ref{fig:sample_by_iter_prob} highlights once more the large
discrepancy between problem instance classes: while for some classes
any optimization algorithm quickly determines a quantum state that has
high probability of yielding the optimal string, other classes of
problems appear out of reach, especially for the local optimization
algorithms. Global optimization looks more promising, but it is
affected by a different set of issues which may limit its practical
usefulness. We remark that optimizing the expected objective function
value of the binary strings may not be the best possible approach, if
the end goal is simply to reach a certain probability of sampling an
optimum (rather than aiming for the ground state, that has probability
1 of sampling an optimum).

\subsection{Entanglement vs.\ no entanglement}
\label{s:entnoent}
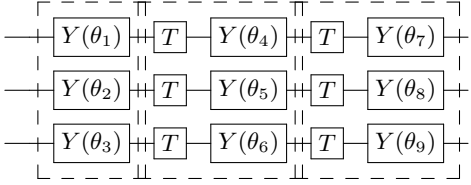
\begin{figure}
 \leavevmode
 \centering
 \Qcircuit @C=1em @R=.6em {
   & \qw & \gate{Y(\theta_1)} & \gate{T} & \gate{Y(\theta_4)} & \gate{T} & \gate{Y(\theta_7)} & \qw\\
   & \qw & \gate{Y(\theta_2)} & \gate{T} & \gate{Y(\theta_5)} & \gate{T} & \gate{Y(\theta_8)} & \qw\\
   & \qw & \gate{Y(\theta_3)} & \gate{T} & \gate{Y(\theta_6)} & \gate{T} & \gate{Y(\theta_9)} & \qw
   \gategroup{1}{3}{3}{3}{1.3em}{--} \gategroup{1}{4}{3}{5}{1.3em}{--}
   \gategroup{1}{6}{3}{7}{1.3em}{--}
 }
  \caption{Example of the variational form without entanglement on three qubits. Each box represents a layer.}
  \label{fig:var_circuit_phase}
\end{figure}

The variational form used throughout the paper introduces entanglement
after the first layer of Y rotations. Experiments discussed in Section
\ref{s:conviter} do not show any clear advantage for the variational
forms 2L and 3L that use entanglers, as compared to 1L that generates
product states only. Indeed, Fig.~\ref{fig:conv_by_iter} reports a
marginal improvement with 2L and 3L using COBYLA, but this comes at
the cost of several additional iterations of the optimization
algorithm (recall that the $x$-axis is normalized by $q\ell + 1$,
where $\ell$ is the number of layers); with RBFOpt, the variational
form 1L achieves the best results. Other local solvers yield results
consistent with COBYLA. We now try to understand whether 2L, 3L can
truly improve performance of the local solvers because of
entanglement, or if the reason for such improvement could be
attributed to other factors, e.g., better chance to escape local
minima. To do so, we repeat the experiments and the analysis using a
variational form that mimicks the one described in
Sec.~\ref{s:varform}, but does not use two-qubit gates, i.e., the CZ
gates of Fig.~\ref{fig:var_circuit}. Having multiple adjacent Y
rotations on the same qubit would amount to introducing copies of the
same variational parameter, which is undesirable from an optimization
standpoint. Hence, after each layer of Y rotations we apply a T gate
on each qubit. This way, each variational parameter on the same qubit
has a different effect. We obtain a variational form exemplified in
Fig.~\ref{fig:var_circuit_phase}.

\begin{figure}[tb]
  \subfloat[Constrained Optimization By Linear Approximation (COBYLA).]{
    \includegraphics[width=0.5\textwidth]{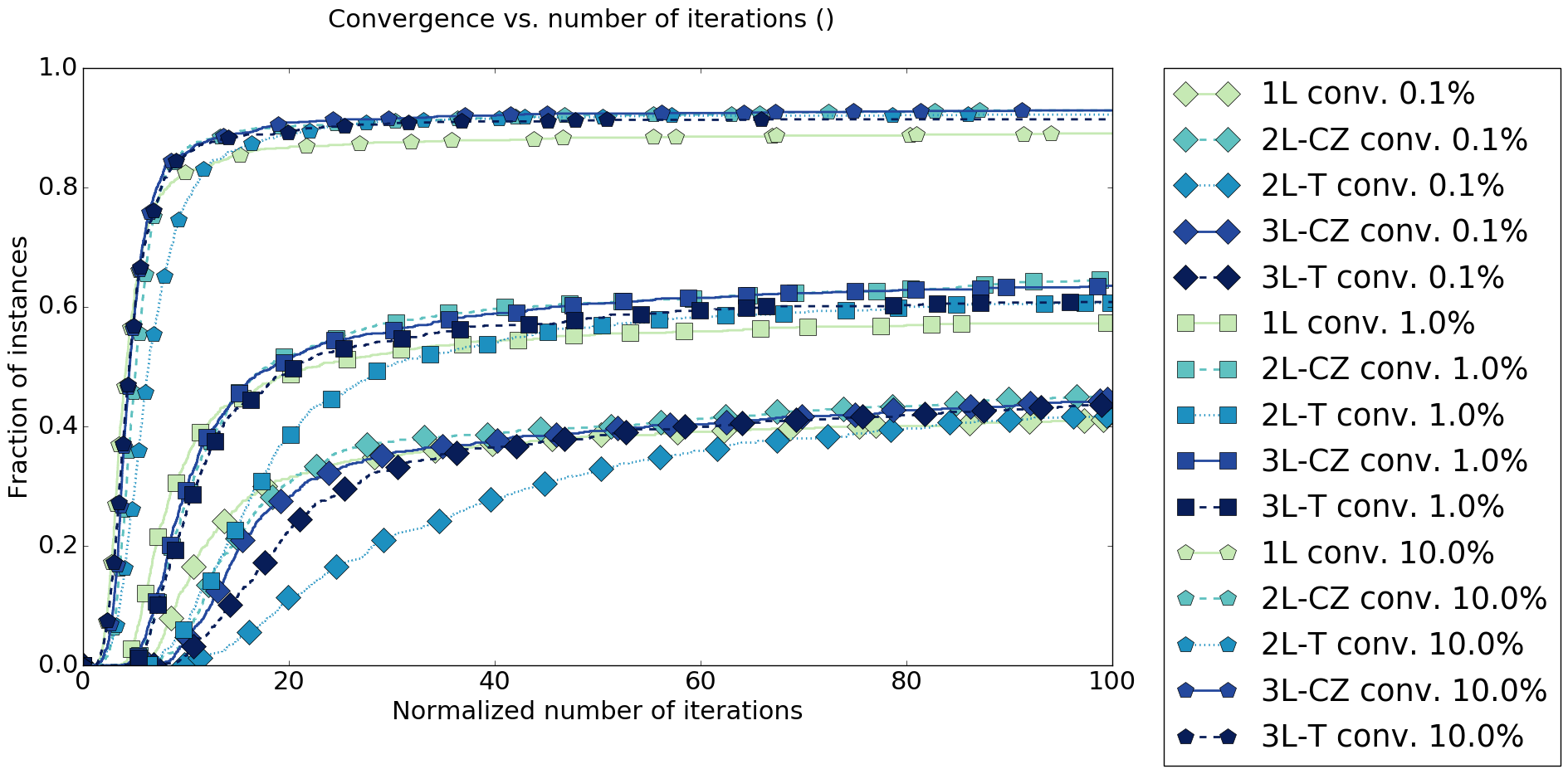}
  }

  \subfloat[RBFOpt.]{
    \includegraphics[width=0.5\textwidth]{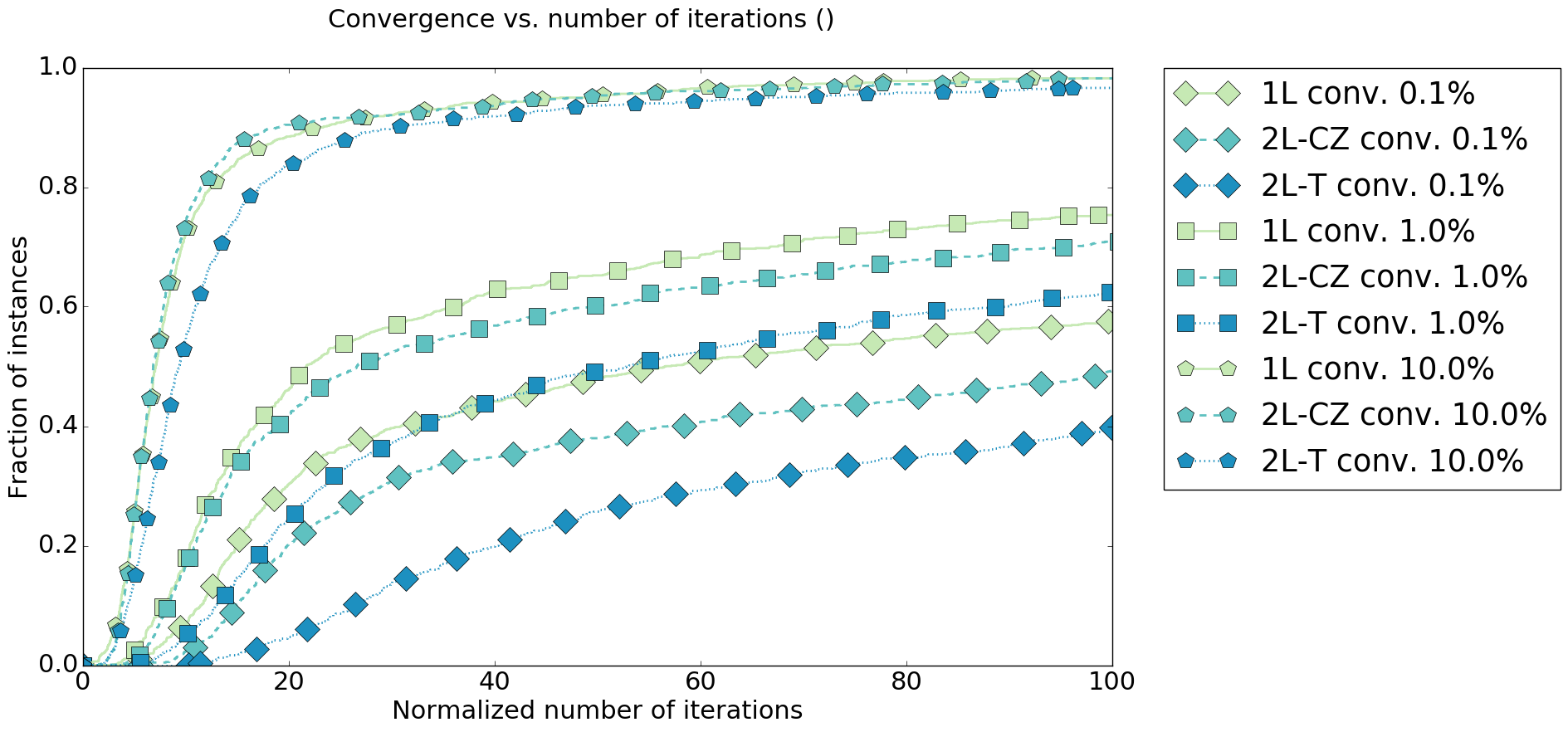}
  }
  
  \caption{Fraction of the instances on which a given algorithm
    converges to the specified tolerance, versus the normalized number
    of iterations. We compare variational forms with and without
    entanglement.}
  \label{fig:conv_by_iter_phase}
\end{figure}

\begin{figure}[tb!]
  \subfloat[Constrained Optimization By Linear Approximation (COBYLA).]{
    \includegraphics[width=0.5\textwidth]{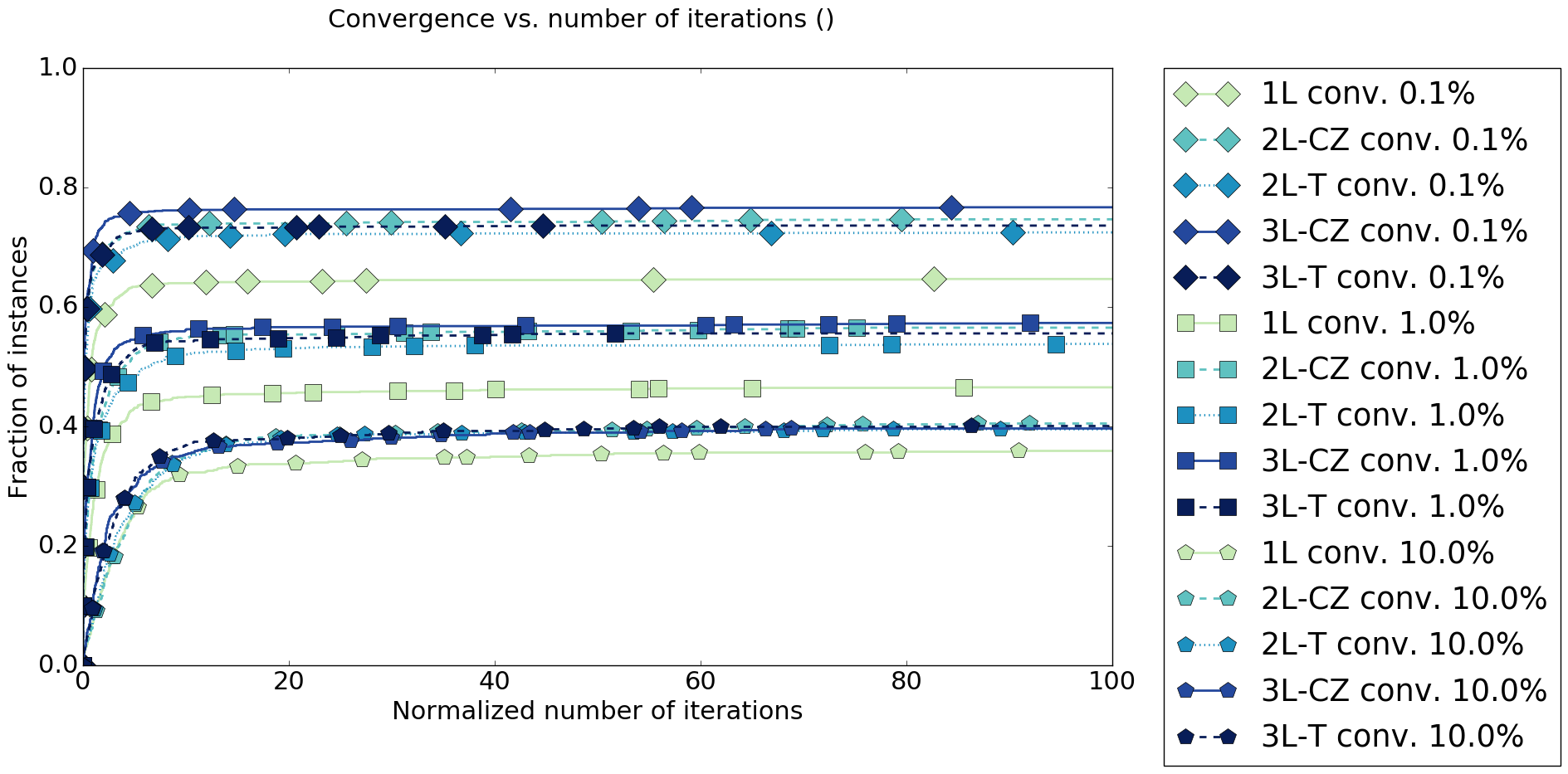}
  }

  \subfloat[RBFOpt.]{
    \includegraphics[width=0.5\textwidth]{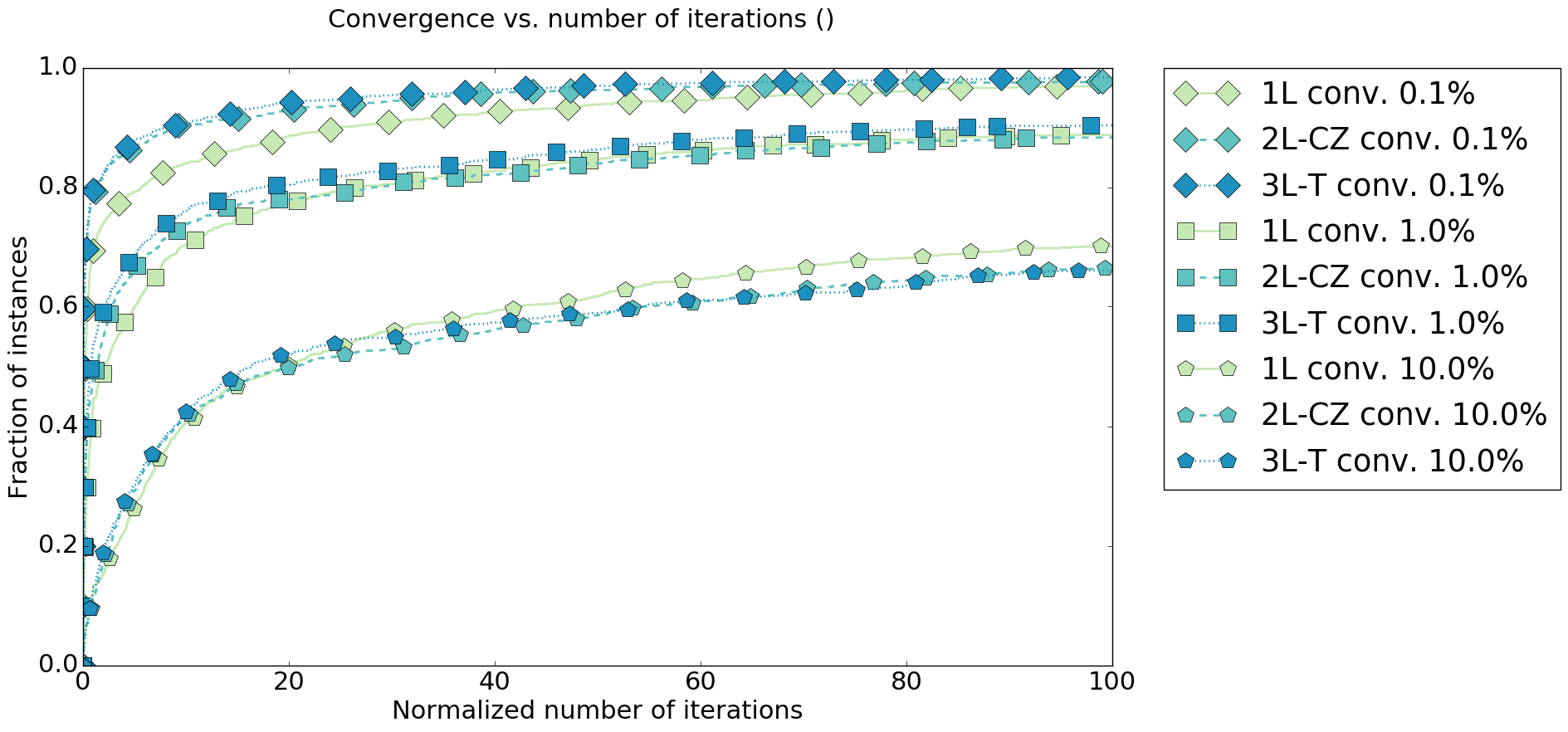}
  }
  
  \caption{Fraction of the instances on which a given algorithm
    explores at least one quantum state with probability of sampling
    the optimal solution greater than a given threshold, versus the
    normalized number of iterations. We compare variational forms with
    and without entanglement.}
  \label{fig:sample_by_iter_phase}
\end{figure}

We compare the performance of the optimization algorithms using the
previous variational form (which we label 2L-CZ, 3L-CZ) and the new
variational form without entanglement (labeled 2L-T, 3L-T). A summary
of the results is given in Fig.~\ref{fig:conv_by_iter_phase},
reporting convergence versus number of iterations as in Section
\ref{s:conviter}, and Fig.~\ref{fig:sample_by_iter_phase}, reporting
probability of sampling an optimal solution versus number of
iterations as in Section \ref{s:sampling}.
Fig.~\ref{fig:conv_by_iter_phase} shows that for COBYLA, the
performance of 2L-CZ, 2L-T, 3L-CZ, 3L-T is essentially
indistinguishable. This is confirmed by
Fig.~\ref{fig:sample_by_iter_phase}, looking at the probability of
sampling an optimal solution. The variational form with two and three
layers still appear to be better than 1L, but there does not seem to
be any significant difference between using CZ gates or T gates. A
possible explanation is that the local optimization algorithm finds
better solutions (in a larger number of iterations) when there are
additional variational parameters that can be used to avoid being
trapped in a local minimum. Similar results are obtained with LBFGS.
The results with RBFOpt tell a different story: here 2L-CZ is
sinificantly better than 2L-T in Fig.~\ref{fig:conv_by_iter_phase},
even though attaining a better objective function value does not seem
to affect the probability of sampling an optimal solution, see
Fig.~\ref{fig:sample_by_iter_phase}. The variational form 1L is still
better than any 2L variational form with RBFOpt, as remarked in
Section \ref{s:conviter}.

The results reported in this section paint a mixed picture. They
suggest that there may be nothing special about variational form with
entangling CZ gates, because we can achieve similar results with a
different variational form that does not introduce entanglement. The
variation in performance of the various algorithm may be attributed to
characteristics of the algorithms themselves, rather than of the
variational form: some algorithms seem to benefit from having
additional variational parameters to optimize, while others do
not. Overall, while the experiments are not conclusive, we are unable
to observe any advantage in using two-qubit gates in our setting, as
compared to variational forms with single-qubit gates only.

\section{Conclusions}
\label{s:conclusions}
Our study of VQE on classical combinatorial optimization problems
highlights several obstacles that need to be overcome in order to make
it a viable method with potential to be used in practice. The first
obstacle is that local optimization algorithms that aim at first-order
stationary points appear to get stuck in local minima very frequently
on certain problem classes. A possible solution is to use different
classical optimization algorithms, more suited for nonconvex problems,
or a different variational form. A second obstacle is that convergence
to the optimum is slow, requiring more and more iterations of the
classical optimization routine as the problem size increases. This
difficulty may be alleviated by computing derivatives on the quantum
computer, rather than relying on derivative-free optimization methods
or finite-difference estimations of the gradient; the computation of
derivatives is possible for certain types of variational forms. A
different choice of variational form may also help. The third obstacle
is that the performance of VQE on classes of problems correlates with
the performance of a classical Branch-and-Bound solver on a naive
translation of the Hamiltonian to a quadratic form over binary
variables: since VQE performs well only on problems for which the
classical Branch-and-Bound also performs well, there is little
advantage to be gained. This is likely due to the fact that these hard
problems have dense quadratic matrices in the classical
representation, and the corresponding Hamiltonians have many distinct
eigenvalues. A possible way forward is to start exploiting problem
structure in a systematic way, moving away from the problem-agnostic
nature of VQE; this may also alleviate some of the other issues.

Our experiments indicate that there does not seem to be a significant
gain in performance, if any, by using two-qubit gates in the
variational form, as compared to single-qubit gates. Of course, if the
variational form yields a product state the computation could be
performed efficiently on a classical computer, hence suggesting that
VQE does not yield any quantum speedup on this class of
problems. However, two important remarks are in order: first, binary
optimization problems by construction admit a ground state that is a
basis state, therefore they are very poor candidates to showcase the
benefits of entanglement; second, from a theoretical point of view we
know that two-qubit gates can be useful even for binary optimization
problems (e.g., \cite{farhi2014quantum} shows that we can essentially
simulate adiabatic optimization with a problem-dependent variational
form with a sufficient number of gates). Hence, while two-qubit gates
do not seem to yield benefits in the setting of this paper, this
conclusion would change on a different class of problems or with a
different optimization approach than a problem-agnostic VQE.

Ultimately, the most important question is to understand whether a
VQE-like approach has potential to be competitive with classical
combinatorial optimization methodologies. At the scale at which we are
able to simulate, which is approximately the scale of existing
superconducting qubit devices, this question cannot be answered:
algorithms and software for binary optimization on classical computers
are very refined, and optimal solutions for the problems discussed in
this paper can typically be found in fractions of a
second \footnote{The times reported in Fig.\ref{fig:cplex_times} are
  largely due to the effort of {\it proving} optimality of the
  solution, rather than {\it finding} the solution.}. On the other
hand, the VQE implementation tested in this paper requires the
exploration of hundreds or thousands of trial states as well as
iterations of a classical optimization algorithm), and may fail to
converge anyway. Leaving aside considerations on hardware efficiency,
this suggests that the performance of VQE must be greatly increased
before it can be considered competitive. Our study provides some
suggestions on possible directions of improvement.

\bibliography{quantum,phd}

\clearpage
\appendix

\end{document}